\documentclass{aa}  
\usepackage{graphicx}
\usepackage{txfonts}
\usepackage{lipsum}
\usepackage{subcaption}         
\usepackage{lscape}             
\usepackage{placeins}           
 \usepackage{multirow}
\usepackage{threeparttable}
\newcommand{\degree}{\ensuremath{^\circ}\xspace}
\newcommand{\source}{4U~1822--37}
\usepackage{hyperref}

\begin{document}

   \title{The first IXPE view of the eclipsing accretion-disc corona source 4U 1822--37}


\author{A. Anitra\inst{\ref{in:Uni_Palermo}}\fnmsep\thanks{alessio.anitra@inaf.it}
    \and A. Gnarini\inst{\ref{in:MSFC}}
    \and T. Di Salvo\inst{\ref{in:Uni_Palermo}}
    \and R. Iaria \inst{\ref{in:Uni_Palermo}}
    \and A. Marino \inst{\ref{in:USACH},\ref{in:CIRAS}, \ref{in:ICE-CSIC},\ref{in:IEEC},\ref{in:IASF}}
    \and F. Barra  \inst{\ref{in:IASF},\ref{in:Uni_Palermo}}
    \and L. Burderi \inst{\ref{in:Uni_Cagliari},\ref{in:IASF}}
    \and A. Sanna \inst{\ref{in:Uni_Cagliari}}
    \and L. Marra\inst{\ref{in:INAF-IAPS}}
    \and S. Bianchi\inst{\ref{in:UniRoma3}}
    \and G. Matt\inst{\ref{in:UniRoma3}}
    \and F. Ursini\inst{\ref{in:UniRoma3}}
    \and C. Miceli\inst{\ref{in:Uni_barcelona}}
    \and A. Miraval Zanon\inst{\ref{in:Asi}}
    \and W. Leone \inst{\ref{in:Uni_Cagliari}}
    \and F. Capitanio\inst{\ref{in:INAF-IAPS}}
    \and S. Fabiani\inst{\ref{in:INAF-IAPS}}
    \and P. Kaaret\inst{\ref{in:MSFC}}
    \and A. Tarana\inst{\ref{in:INAF-IAPS}}
    }

\institute{Dipartimento di Fisica e Chimica - Emilio Segrè,
Universit\`a di Palermo, via Archirafi 36 - 90123 Palermo, Italy \label{in:Uni_Palermo}
    \and NASA Marshall Space Flight Center, Huntsville, AL 35812, USA \label{in:MSFC}
    \and Departamento de Física, Universidad de Santiago de Chile (USACH), Av. Víctor Jara 3493, Estación Central, Chile\label{in:USACH}
    \and
    Center for Interdisciplinary Research in Astrophysics and Space Sciences (CIRAS), Universidad de Santiago de Chile.\label{in:CIRAS}
    \and
    Institute of Space Sciences (ICE, CSIC), Campus UAB, Carrer de Can Magrans s/n, E-08193 Barcelona, Spain\label{in:ICE-CSIC}
    \and
    Institut d'Estudis Espacials de Catalunya (IEEC), 08860 Castelldefels (Barcelona), Spain\label{in:IEEC}
    \and
    INAF - IASF Palermo, via Ugo La Malfa 153, I-90146 - Palermo, Italy\label{in:IASF}   
    \and Dipartimento di Fisica, Universit\`a degli Studi di Cagliari, SP Monserrato-Sestu, KM 0.7, Monserrato, 09042 Italy \label{in:Uni_Cagliari}
    \and  
     INAF -- Istituto di Astrofisica e Planetologia Spaziali, Via del Fosso del Cavaliere 100, 00133 Roma, Italy \label{in:INAF-IAPS}
    \and Dipartimento di Matematica e Fisica, Università degli Studi Roma Tre, via della Vasca Navale 84, I-00146 Roma, Italy \label{in:UniRoma3}
    \and Universitat Politècnica de Catalunya,Departament de Física, EEBE , Av. Eduard Maristany 16, 08019, Barcelona, Spain \label{in:Uni_barcelona}
    \and  ASI -- Agenzia Spaziale Italiana, Via del Politecnico snc, I-00133 Rome, Italy \label{in:Asi}
    }

\date{Received September 30, 20XX}

\abstract
{4U~1822--37 is a high-inclination eclipsing neutron-star low-mass X-ray binary and a prototype accretion-disc corona (ADC) source, in which the direct inner emission is largely obscured and the observed X-ray flux is dominated by scattered radiation. X-ray polarimetry provides a direct probe of this scattering geometry.}
{We investigated the geometry of 4U~1822--37 through its first X-ray spectro-polarimetric observation, testing the high-inclination ADC scenario with simultaneous broadband spectroscopy and polarimetry.}
{We analysed coordinated \textit{IXPE}, \textit{XMM-Newton}, \textit{NuSTAR}, and \textit{Swift} observations. We performed broadband spectral modelling and model-independent and spectro-polarimetric analyses of the \textit{IXPE} data, including energy and orbital phase-resolved polarimetry.}
{The broadband spectrum is described by a soft thermal component, a Comptonised continuum, a hard power-law tail, and relativistically blurred reflection. The observed luminosity, $L_{\rm obs}\simeq6.1\times10^{36}\ {\rm erg s^{-1}}$, is well below the intrinsic luminosity expected from the orbital and spin evolution, consistent with the presence of an extended, optically thin corona that scatters only a fraction of the intrinsic radiation into the line of sight. In the 2--8 keV band, \textit{IXPE} measures a polarisation degree (PD) of $7.9\pm0.6\%$ and a polarisation angle (PA) of $-24^\circ\pm2^\circ$. The PD increases with energy, while the PA remains approximately constant. Spectro-polarimetric modelling confirms the energy trend, although the relative contributions of the Comptonised continuum and reflection remain degenerate. The orbital-phase-resolved analysis suggests a lower PD during an eclipse, ${\rm PD}=5.5\pm1.7\%$, with no significant PA change or major broadband-continuum variation above 2 keV.}
{The high polarisation degree, the stable polarisation angle, the increase in PD with energy, and the decrease in PD during an eclipse all support the view that 4U~1822--37 is observed in an extreme high-inclination, scattering-dominated regime. In this picture, the extended corona plays a central role in shaping the observed X-ray emission and its polarisation properties.}


\maketitle
 \nolinenumbers
\section{Introduction}
The accretion flow around neutron stars (NSs) in low-mass X-ray binaries (LMXBs) \citep[see e.g.][for a review]{DiSalvo_2023hxga.book..147D} typically includes an optically thick accretion disc, a boundary layer connecting the disc to the NS surface, and a cloud of hot electrons (the ‘corona’). However, when these systems are viewed at very high inclinations, the vertically extended outer rim of the disc can block a direct view of radiation from the innermost regions. In these sources, the continuum reaching the observer is mostly produced by electron scattering and reprocessing in an extended, photoionised medium above the disc, commonly interpreted as a sandwich-shaped corona \citep[e.g. ][]{Church_2001AdSpR..28..323C} and/or an equatorial thermal-radiative wind \citep[e.g. ][]{1983ApJ...271...70B}. For this reason, such systems are often referred to as accretion disc corona (ADC) sources.
Among ADC sources, 4U~1822--37 stands out as one of the most puzzling systems. It is an eclipsing LMXB discovered by \cite{1822_1978Natur.276..247G} with an orbital period of 5.57~h \citep{jonker_01, Anitra_2021A&A...654A.160A}, hosting a NS that shows periodic X-ray pulsations with a period of 0.59~s \citep[e.g. ][]{Iaria_2024A&A...683A..79I}. As estimated from infrared observations \citep{mason_82} and later confirmed by spectral analyses \citep{Anitra_2021A&A...654A.160A}, the inclination angle to our line of sight lies between 80\degree and 85\degree.

Despite extensive studies, key aspects of {\source}, such as its geometry, luminosity, and spectral features, remain debated. 
As stated by \citet{burderi_10}, the large orbital period derivative, $\dot{P}=1.426(26)\times10^{-10}\ \mathrm{s\,s^{-1}}$, cannot be explained by a conservative mass transfer at the rate implied by the observed luminosity ($\sim10^{36}\ \mathrm{erg\,s^{-1}}$; \citealt{Iaria_2024A&A...683A..79I}). Highly non-conservative mass transfer is required, up to 7 times the Eddington accretion rate for a $1.4\,M_\odot$ NS. In this scenario, the intrinsic luminosity is likely near the Eddington limit ($\sim10^{38}\ \mathrm{erg\,s^{-1}}$), i.e. about two orders of magnitude higher than the observed one.
Moreover, assuming a near-Eddington luminosity implies a magnetic field of $\sim8\times10^{10}\ \mathrm{G}$ \citep{jonker_01}, consistent with the value reported by \citet{Iaria_2024A&A...683A..79I} through the detection of a cyclotron line at $0.66$~keV.

The broadband X-ray spectrum of {\source} is well described by a thermal disc component, a Comptonised component from a compact, optically thick region close to the NS, and a disc-reflection component. In addition, several narrow emission features are observed, including low-energy lines from highly ionised species and narrow Fe-K lines \citep{Iaria_13}. Recently, through X-ray Doppler tomography, \cite{Sameshima_2026PASJ...78..976S} showed that the narrow Fe K$\alpha$ fluorescence line is likely associated with a localised region near the accretion stream--disc overflow, rather than with an axisymmetric disc, the extended corona, the neutron-star surface, or the companion star.

The mismatch between observed and intrinsic luminosity has been addressed with a peculiar geometry. To account for the observed flux at such a high inclination, \citet{Iaria_13} and \citet{Anitra_2021A&A...654A.160A}, and more recently \citet{Iaria_2024A&A...683A..79I}, proposed that even if the central emission is hidden from direct view by the outer disc rim, all the spectral components can still be observed because an extended, optically thin corona surrounding the source (with optical depth $\tau\approx0.05$) scatters a fraction of the radiation into the line of sight. 

Testing such a configuration is challenging when relying on spectroscopy and timing alone, as different geometries can yield similar spectral components and variability.
 Recent advances in X-ray polarimetry, particularly with the Imaging X-ray Polarimetry Explorer (\textit{IXPE}), now enable energy-resolved polarimetry and direct tests of geometrical models \citep{Weisskopf_2022JATIS...8b6002W}. Measuring the polarisation degree (PD) and angle (PA) is crucial for disentangling geometry and radiative mechanisms, because different spectral components (Comptonisation, blackbody emission, and reflection) contribute differently to the net polarisation  \citep{Poutanen2024Galax..12...46P, Fabiani2024A&A...684A.137F}.
 
 The polarisation signal is therefore key to distinguishing among coronal models and shapes, since it is strongly affected by the geometry of the Comptonising region. As shown by \citet{Gnarini.etAl.2022}, both the PD and PA vary significantly between slab-like and spherical coronae, with the slab geometry typically showing a PD that increases with energy. 
 Early \textit{IXPE} observations of typical weakly magnetised NS-LMXBs generally reported low PDs, of the order of \(\sim1\)--2\%
 (see \citealt{Ursini2024Galax..12...43U}). 
 However, ADC sources are expected to show higher PDs because their observed emission is dominated by scattered and reprocessed radiation, as recently supported by the high PD measured with IXPE in 2S 0921--630 \citep{Tomaru_2026MNRAS.547ag498T}.
 In addition, the presence of an extended, optically thin corona in the {\source} should significantly increase the observed PD. The primary interaction in the corona is Thomson scattering, so the scattered radiation is polarised even if the incident radiation is not; the degree of polarisation depends on the viewing angle and is expected to be enhanced in high-inclination configurations \citep{Gnarini.etAl.2022}.

Here we present the first measurement of the X-ray polarisation of 4U~1822--37 obtained with
\textit{IXPE}, reporting a time-averaged polarisation of ${\rm PD}=7.9\pm0.6\%$ in the 2--8 keV band,
with marginal evidence for an increase in PD with energy. We investigate the system geometry
through a combined spectral and polarimetric analysis of X-ray data from a campaign of
simultaneous \textit{XMM-Newton}, \textit{Swift}, \textit{NuSTAR}, and \textit{IXPE}
observations.

\section{Observations and data analysis}
The data analysed in this work are part of the \textit{IXPE} GO Cycle-2 programme [ID 2154; PI: Alessio Anitra], a coordinated campaign that we organised to target \source. The \textit{IXPE} observations were complemented by contemporaneous \textit{NuSTAR}, \textit{XMM-Newton} and \textit{Swift-XRT} pointing obtained under our companion proposals.

\subsection{IXPE}

\textit{IXPE} observed {\source} from 3 October 2025 at 09:24:16 UT to 13 October 2025 at 19:07:26 UT (ObsID: 04003101), with a total net exposure time of 496~ks for each detector unit (DU). We considered a circular region with a 100\arcsec\ radius and an annular region with an inner and outer radius of 180\arcsec\ and 240\arcsec, respectively. Both regions are centred on the source. The radius of the source region was derived using an iterative way to maximise the signal-to-noise ratio (S/N) in the 2--8 keV band (see also \citealt{Piconcelli2004, Ursini2023}). Since the source is faint ($< 1$ count~s$^{-1}$~arcmin$^{-2}$), we performed both background rejection and subtraction \citep{DiMarco2023}. The ancillary response files (ARFs) and the modulation response files (MRFs) were computed using the \texttt{ixpecalcarf} tool, considering the same extraction radius of the source region. Each Stokes $I$ spectrum was rebinned with the \texttt{ftgrouppha} task, requiring a S/N of 3 for each bin. The Stokes $Q$ and $U$ spectra were rebinned considering the same grouping scheme as the $I$ spectra. We computed the light curves for each DU employing the \texttt{extractor} tool, considering 1000s time bins. 

\subsection{NuSTAR}

The Nuclear Spectroscopic Telescope Array (\textit{NuSTAR}) observed the source on 10 October 2025 (ObsID 31101003002) for a total exposure of 83~ks. Data reduction and the extraction of spectra and light curves were performed with \texttt{nupipeline} and \texttt{nuproducts} within HEASoft, using the latest CALDB version available at the time of the analysis (\texttt{indx20251027}). Source events were extracted from a circular region with a radius of 100\arcsec, while background events were taken from an equally sized circular region on the same detector quadrant.

We grouped the spectra of the two focal-plane modules, FPMA and FPMB, using \texttt{ftgrouppha}, applying optimal rebinning \citep{Kaastra_16} and requiring at least 25 counts per energy bin. Barycentric corrections were not applied to the products used for the time-averaged spectral analysis presented here; a dedicated timing analysis, including an updated orbital ephemeris, is beyond the scope of this work and will be addressed in a future study.

\subsection{XMM-Newton}

The source was observed by \textit{XMM-Newton} on 7 October 2025 (ObsID 0973390501) for 23~ks. During this observation, the \textit{EPIC-pn} and MOS2 cameras operated in Timing mode with the thick filter, while MOS1 was set to Small Window mode with the thick filter.
We reduced the \textit{XMM-Newton} data with the Science Analysis Software (SAS) v22.1.0. First, we extracted the high-energy ($10$--$12$~keV) single-event light curve to identify intervals of flaring particle background. The light curve shows strong flaring after the first $\sim16$~ks from the start of the observation. To mitigate the effect of flares, we selected good time intervals with count-rate thresholds of $<0.95$~cts~s$^{-1}$ for \textit{EPIC-pn} and $<0.1$~cts~s$^{-1}$ for the two MOS cameras. The flare-cleaned pn and MOS source light curves exhibit a partial eclipse around $9000$~s from the start of the observation.
As shown in previous works \citep{Iaria_13, Anitra_2021A&A...654A.160A}, the broadband continuum extracted during eclipse does not differ strongly from the phase-averaged spectrum. We therefore did not exclude eclipse intervals from the time-averaged spectro-polarimetric analysis, although we later used eclipse-selected spectra to check the consistency of the polarimetric results.

Although the source is not particularly bright, with an average $0.5$--$12$~keV \textit{EPIC-pn}/MOS count rate of $\sim40$~cts~s$^{-1}$ in the source region, we verified the pile-up fraction with the SAS tool \texttt{epatplot} for \textit{EPIC-pn} and both MOS observations. \textit{EPIC-pn} shows observed-to-model single and double event fractions consistent with unity across $0.5$--$2.0$~keV. Conversely, both MOS cameras show single-fraction values below unity and double-fraction values significantly above the expected distribution, indicating pile-up.
We attempted to mitigate pile-up in the MOS data by excluding the central CCD columns, where the source is brightest. However, this led to poorer statistics and to a background-dominated spectrum above $\sim7$~keV. Since the same energy range is covered by \textit{EPIC-pn}, which offers better spectral resolution and effective area and is not affected by pile-up, we retained only the \textit{EPIC-pn} data for the spectral analysis and did not use MOS.

The \textit{EPIC-pn} source and background spectra were extracted from RAWX columns 28--48 and 3--6, respectively. The total useful \textit{EPIC-pn} exposure after filtering is 12~ks. As for the \textit{NuSTAR} spectra, we grouped the spectrum with \texttt{ftgrouppha}, using optimal rebinning (\texttt{grouptype=optmin}) and requiring at least 25 counts per energy bin \citep{Kaastra_16}. According to the SAS Data Analysis thread\footnote{\url{https://www.cosmos.esa.int/web/xmm-newton/sas-thread-evenergyshift}}, bright sources observed in \textit{EPIC-pn} Timing mode may exhibit count-rate-dependent energy-scale shifts. We applied the recommended correction following the SAS \texttt{evenergyshift} thread and re-extracted the spectra; details are reported in Appendix~\ref{epic_shift}.

Reflection Grating Spectrometer (\textit{RGS}) data were processed with the standard \texttt{rgsproc} pipeline up to the spectra stage. After \texttt{rgsproc}, we checked the source list to ensure that the target marked as \texttt{PROPOSAL} was the prime source and that its coordinates were correct. We verified the extraction masks by inspecting the dispersion--cross-dispersion and dispersion--energy images, and confirmed that the background regions did not intersect any bright sources. To mitigate high particle background, we built background light curves from CCD~9 for \textit{RGS}~1 and \textit{RGS}~2 and created additional good time intervals (GTIs) by selecting intervals with rates below 0.1~cts~s$^{-1}$. First- and second-order spectra and response matrices were produced by \texttt{rgsproc}. After inspecting the second-order products, we discarded them because they were strongly background-dominated. Finally, we combined the first-order spectra, responses, and backgrounds from \textit{RGS}~1 and \textit{RGS}~2 with \texttt{rgscombine}.

\subsection{Swift XRT}
The X-ray telescope (XRT) on board the Neil Gehrels \textit{Swift} observatory \citep{Gehrels2004ApJ...611.1005G} observed 4U~1822--37 five times in 2025, between 3 and 11 October 2025, with an intra-observational cadence of two days. Each observation was performed in Window Timing (WT; readout time of 1.77\,ms) mode and lasted for about 2~ks. Data were processed with the \texttt{xrtpipeline} task, adopting standard cleaning criteria. For the spectral analysis, we selected events with grades 0--12. A circular region with a radius of 47 arcseconds centred on the source coordinates was adopted as the source region. In contrast, a region of the same size and shape located far from the source position was used to estimate the background. In all observations, we evaluated the impact of potential pile-up to be negligible. We generated the spectra and light curves with the \texttt{xrtproducts} task. The \textit{Swift}/XRT background-subtracted spectra were grouped with \texttt{ftgrouppha}, applying optimal rebinning \citep{Kaastra_16} and requiring at least 25 counts per energy bin, consistently with the grouping adopted for the \textit{NuSTAR} and \textit{XMM-Newton} spectra.

\section{Spectral analysis}

We first performed a simultaneous fit to characterise the broadband continuum and to verify consistency among the instruments. The adopted energy ranges were 0.5--2 keV for \textit{RGS}, 2--10 keV for \textit{EPIC-pn}, 4.5--35 keV for \textit{NuSTAR}, and 2--9 keV for the five \textit{Swift/XRT} spectra. 
The \textit{EPIC-pn} data below 2 keV were excluded because the soft band is affected by known cross-calibration differences between instruments \citep{Kirsch2004SPIE.5488..103K}; since \textit{RGS} is available and provides both more reliable calibration and higher resolving power in this range, we relied on it at low energies. 
For \textit{NuSTAR}, we excluded channels below 4.5 keV, where a known excess in FPMA produces systematic deviations at the lowest energies \citep{Madsen_2020arXiv200500569M}, and above 35 keV, where the background dominates for this source. 
The \textit{Swift/XRT} spectra were limited to 2--9 keV for the same practical reasons.

The broadband spectral analysis presented here is based on phase-averaged spectra. This choice is motivated by previous phase-resolved studies of {\source}, which showed that the main continuum components remain relatively stable along the orbit \citep{Iaria_13}, while the strongest phase-dependent changes are mostly associated with emission lines. Therefore, although this approach is not intended to describe the detailed orbital evolution of the emitting and reprocessing material, it provides a suitable broadband baseline for the spectro-polarimetric analysis presented in this work.

Spectral fitting was performed with \textsc{XSPEC} v12.15.0d \citep{Arnaud_1996}. 
The continuum was modelled with a thermal blackbody (\texttt{bbodyrad}) and a Comptonised component obtained by convolving a blackbody seed spectrum with \texttt{thcomp} \citep{Zdziarski_2020MNRAS.492.5234Z}. 
Interstellar absorption was taken into account using the Tuebingen-Boulder ISM absorption model \texttt{TBabs}, adopting the abundances of \citet{Wilms_2000ApJ...542..914W} and the cross sections of \citet{Verner_96}. 
A multiplicative constant was included to account for inter-instrument cross-normalisation differences. It was fixed to unity for the reference dataset (\textit{NuSTAR}/FPMA) and left free for all the others. We verified that the best-fit values of the cross-normalisation constants were always between 0.7 and 1.3.

\begin{figure}
    \centering
    \includegraphics[width=0.5\textwidth]{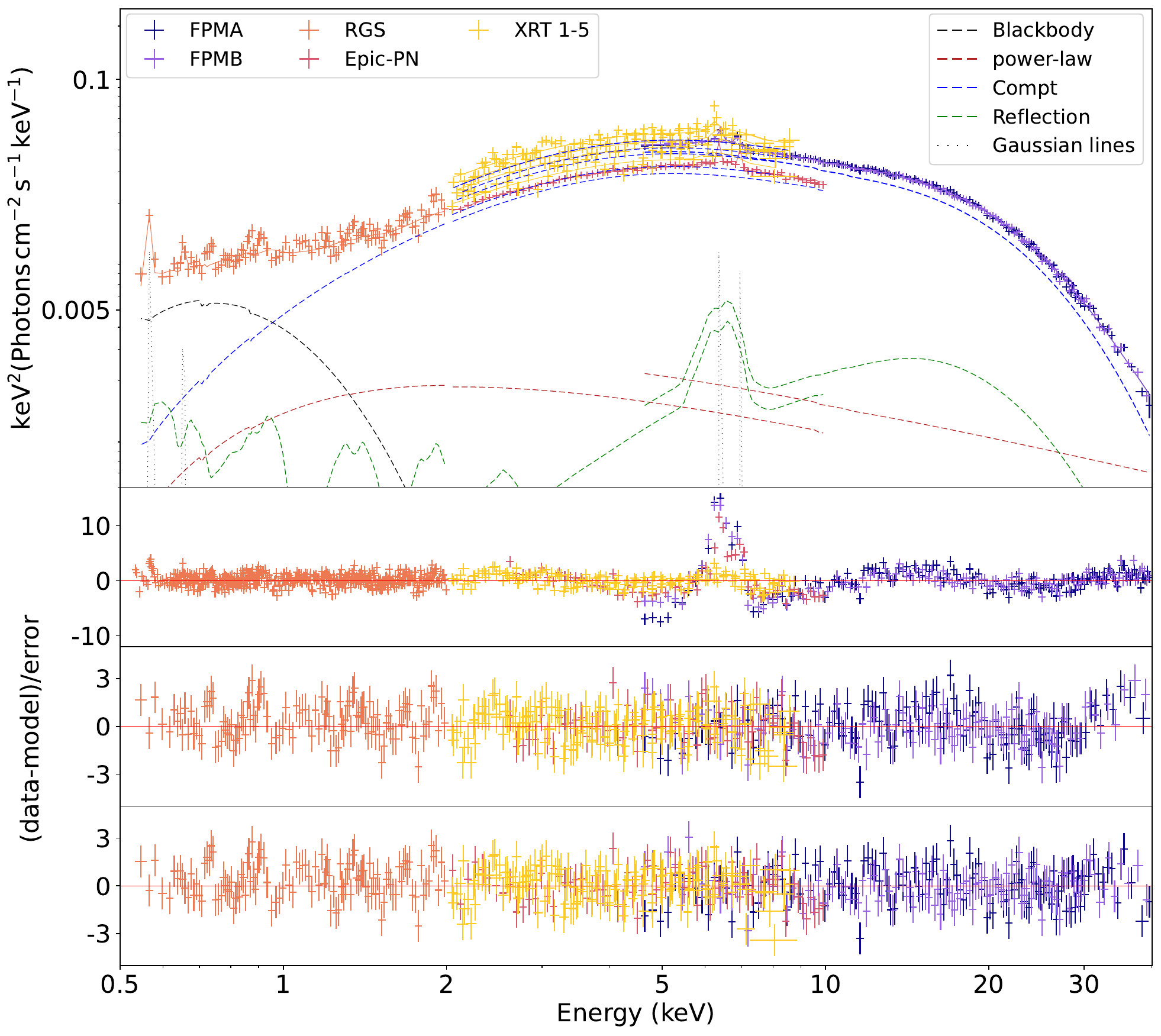}
    \caption{Unfolded broadband spectrum obtained from \texttt{Model 2}. The lower panels show the residuals, in units of $\sigma$, for the main fitting steps discussed in the text: the baseline continuum model before adding the Gaussian emission lines, the model including 5 Gaussian lines (\texttt{Model 1}), and the final reflection model \texttt{Model 2}.}
    \label{fig:reflection_plot}
\end{figure}

With this baseline model, the joint fit revealed narrow residuals in the \textit{RGS} band consistent with O\,VII and O\,VIII emission lines, a structured iron complex between 6.4 and 7.0 keV, and, in the \textit{NuSTAR} spectra, a broad excess peaking in the $\sim$10--30 keV range together with absorption-like features near 9.6 and 11~keV (see Fig.~\ref{fig:reflection_plot}). 
We modelled the latter with two absorption edges and improved the soft band by adding two Gaussian emission lines at 0.57 and 0.65~keV. A comprehensive study of the line-emitting plasma in the outer disc or wind, including self-consistent modelling, is beyond the scope of this work and will be presented elsewhere.

To reproduce the Fe K structure, we introduced three Gaussian components to describe emission from neutral iron, Fe\,XXV, and Fe\,XXVI. The resulting phenomenological model (hereafter \texttt{Model 1}) is 
\begin{equation*}
\begin{aligned}
\;& \texttt{Model 1 = TBabs*2edge*(bbodyrad} + \\
& \texttt{thcomp*bbodyrad + 5gauss)} 
\end{aligned}
\end{equation*}
and provides an acceptable fit with $\chi^{2}/{\rm d.o.f.}=1167/983$.
Of the three Gaussian lines in the Fe K band, two are narrow, with centroids at $6.34 \pm 0.01$ keV and $6.87 \pm 0.02$ keV and widths of $\sigma=0.09\pm0.02$ keV, while the third component is much broader, centred at $5.8\pm0.1$ keV with $\sigma=0.9\pm0.1$ keV. Its large width suggests an origin in the innermost disc and points to the presence of an underlying reflection component. As a first-order estimate of the line significance, we considered the normalisation of each Gaussian component and compared it with the corresponding $1\sigma$ uncertainty. With this approximation, all five Gaussian lines are detected above the  $5\sigma$ level. In particular, the two narrow Fe-K lines are detected at a formal significance of about  $20\sigma$.

Inspection of the residuals also reveals a hard excess above $\sim$30 keV (see Fig.~\ref{fig:reflection_plot}). Similar high-energy tails are frequently observed in LMXBs \citep[e.g.][]{Dai_2010A&A...516A..36D} and are often interpreted as an additional Comptonised component produced by a non-thermal electron population \citep{DiSalvo_2000ApJ...544L.119D, DiSalvo2006ApJ...649L..91D}. 
To phenomenologically account for this contribution without affecting the low-energy continuum, we included a power-law component with an exponential low-energy roll-off fixed at the seed-photon temperature of the thermal Comptonisation, i.e. \texttt{expabs*powerlaw}. 
The inclusion of this component significantly improves the fit, yielding $\chi^{2}/{\rm d.o.f.}=1133/982$. An F-test gives a chance improvement probability of $\sim4\times10^{-6}$, corresponding to a significance of $\sim4.5\sigma$. We therefore retained this component in the subsequent reflection modelling.

Motivated by both the large width of the broad Fe K component and the presence of a broad hump in the residuals, we then tested a scenario in which reflection from the accretion disc is included, as first suggested by \citet{Anitra_2021A&A...654A.160A} and more recently supported by \citet{Iaria_2024A&A...683A..79I}.
We replaced the broad Gaussian with a self-consistent reflection component. Specifically, we used \texttt{rfxconv} \citep{Kolehmainen_2011MNRAS.416..311K} to model reflection from an ionised disc, applied to a Comptonised continuum whose parameters were linked to those of the \texttt{thcomp*bbodyrad} component already included in the fit. 
To account for Doppler broadening and relativistic effects in the inner disc, we convolved the reflected spectrum with \texttt{relconv} \citep{Fabian_1989MNRAS.238..729F}. This model assumes a broken power-law emissivity profile with two indices, defined inside and outside a break radius $R_{\rm br}$. Since our spectra are not sufficiently sensitive to constrain distinct emissivity slopes, we tied the two indices together. 
This final reflection model, hereafter \texttt{Model 2}, can be written as
\begin{equation*}
\begin{aligned}
\;&\texttt{Model 2} = \texttt{TBabs*2edge*(4gauss + expabs*powerlaw +} \\
& \texttt{bbodyrad + thcomp*bbodyrad +} \\
& \texttt{relconv*rfxconv*thcomp*bbodyrad})
\end{aligned}
.\end{equation*}
Because the inclination of the source is well constrained in the literature, we fixed it to $82.5^{\circ}$ \citep{Anitra_2021A&A...654A.160A, Iaria_2024A&A...683A..79I}, while leaving the inner disc radius free to vary.  To include only the reflected component, we adopted a negative reflection factor, \texttt{rel\_refl}, and allowed it to vary between $-1$ and $0$.
The addition of the reflection model further improved the fit, yielding $\chi^{2}/{\rm d.o.f.}=1123/979$. The best-fit parameters are reported in Table \ref{reflection_table}, and the unfolded spectrum together with the residuals is shown in Fig.~\ref{fig:reflection_plot}. 
We obtained a relatively strong reflection component, with a reflection fraction of $0.62 \pm 0.03$, an inner disc radius of $59^{+7}_{-9}\,R_{\rm g}$, and a disc ionisation of $\log(\xi /{\rm erg\,cm\,s^{-1}})=2.02 \pm 0.02$. 
The total unabsorbed flux in the 0.1--100 keV band is $1.37\times10^{-9}\ {\rm erg\ s^{-1}\ cm^{-2}}$.
Assuming a distance of 6.1~kpc \citep{Arnason_2021MNRAS.502.5455A}, this corresponds to an unabsorbed luminosity of $6.1\times10^{36}\ {\rm erg\ s^{-1}}$ in the 0.1--100 keV band.

\begin{table}[h!]
\caption{\label{reflection_table}
Best-fit parameters obtained for the two spectral models discussed in the text.}
\centering
\renewcommand{\arraystretch}{1.16}
\resizebox{8.9cm}{!}{
\begin{threeparttable}
\begin{tabular}{llcc}
\hline
\multicolumn{1}{l}{\multirow{2}{*}{Component}} & \multirow{2}{*}{Parameter} & \multicolumn{2}{c}{Model} \\
\multicolumn{1}{c}{} & & {\tt Model 1}\tnote{\#} & {\tt Model 2}\tnote{\#} \\
\hline
{\tt TBabs}
& $N_{\rm H}$ ($10^{22}$ cm$^{-2}$)
& $0.08 \pm 0.04$
& $0.132 \pm 0.006$ \\

{\tt edge}
& E (keV)
& $10.9 \pm 0.2$
& $11.0^{+0.2}_{-0.3}$ \\
& $\tau \, (10^{-2})$ 
& $1.5 \pm 0.5$
& $1.1^{+0.5}_{-0.2}$ \\

{\tt edge}
& E (keV)
& $9.68 \pm 0.09$
& $9.6 \pm 0.2$ \\
& $\tau \, (10^{-2})$ 
& $2.7^{+0.5}_{-0.4}$
& $2.0 \pm 0.4$ \\

{\tt gaussian}
& E (keV)
& $6.34 \pm 0.01$
& $6.324^{+0.023}_{-0.005}$ \\
& $\sigma$ (eV)
& $92 \pm 20$
& $<7$  \\
& Norm($10^{-4}$)
& $(3.8 \pm 0.3)$
& $(2.0 \pm 0.2)$ \\

{\tt gaussian}\tnote{$\ddagger$}
& $E_{\rm line}$ (keV)
& $6.40 \pm 0.02$ 
& $6.39 \pm 0.02$ \\
& $\sigma$ (keV)
& linked\tnote{$\dag$} 
& linked\tnote{$\dag$} \\
& Norm($10^{-4}$)
& $5.2 \pm  0.5$
& $2.9 \pm 0.4 $\\

{\tt gaussian}
& E (keV)
& $5.8 \pm 0.1$
& -- \\
& $\sigma$ (keV)
& $0.9 \pm 0.1$
& -- \\
& Norm($\times10^{-3}$)
& $1.2 \pm 0.2$
& -- \\

{\tt gaussian}
& E (keV)
& $6.87 \pm 0.02$
& $6.91^{+0.09}_{-0.03}$ \\
& $\sigma$ (keV)
& linked\tnote{$\dag$}
& linked\tnote{$\dag$} \\
& Norm($10^{-4}$)
& $2.34 \pm 0.19$
& $9.3^{+1.8}_{-1.5}$ \\

{\tt gaussian}\tnote{$\ddagger$}
& $E_{\rm line}$ (keV)
& $6.94 \pm 0.03$
& $6.98 \pm 0.03$ \\
& $\sigma$ (keV)
& $0.11 \pm   0.02$
& $0.011^{+0.050}_{-0.003}$ \\
& Norm($10^{-4}$)
& $3.7 \pm 0.5$
& $2.1 \pm 0.4$ \\

{\tt bbodyrad}
& $kT_{\rm bb}$ (keV)
& $0.193 \pm 0.013$
& $0.179^{+0.002}_{-0.003}$ \\
& norm ($10^{3}$)
& $1.2 \pm 0.5$
& $1.8^{+1.0}_{-0.1}$ \\

{\sc expabs} 
& $E_{cut}$(keV) 
& - 
& $kT_{\rm bb}$ \\
{\sc powerlaw}  
&Index 
&-
& $1.72^{+0.02}_{-0.04}$ \\
 & N(10$^{-3}$) 
 &-
 & $8.9 \pm 0.6$ \\

{\tt thcomp}
& $\Gamma_{\tau}$
& $1.373 \pm 0.017$
& $1.282^{+0.002}_{-0.005}$ \\
& $kT_{\rm e}$ (keV)
& $4.44 \pm 0.03$
& $4.186 \pm 0.007$ \\
& cov\_frac
& $0.78 \pm 0.03$
& $0.647^{+0.001}_{-0.009}$ \\

{\tt bbodyrad}
& $kT_{\rm bb}$ (keV)
& $1.23 \pm 0.03$
& $1.351^{+0.006}_{-0.003}$ \\
& norm
& $24.1 \pm 2.0$
& $15.94^{+0.04}_{-0.11}$ \\

{\tt relconv}
& Index1
& --
& $2.4^{+0.2}_{-0.1}$ \\
& Incl (deg)
& --
& 82.5\tnote{*} \\
& $R_{\rm in}$ ($R_{\rm g}$)
& --
& $59^{+7}_{-9}$ \\
& $R_{\rm out}$ ($10^{3}\,R_{\rm g}$)
& --
& 1.0\tnote{*} \\

{\tt rfxconv}
& rel\_refl
& --
& $-0.62 \pm 0.03$ \\
& Fe$_{\rm abund}$
& --
& $1.0^{+0.40}_{-0.02}$ \\
& $\log(\xi/{\rm erg\,cm\,s^{-1}})$
& --
& $2.02 \pm 0.02$ \\

{\tt gaussian}
& E (keV)
& $0.653 \pm 0.002$
& $0.654 \pm 0.002$ \\
& $\sigma$ (keV)
& $<0.21$
& $0.002^{+0.003}_{-0.001}$ \\
& Norm ($10^{-4}$)
& $1.0 \pm 0.3$
& $1.3 \pm 0.7$ \\

{\tt gaussian}
&E (keV)
& $0.569 \pm 0.001$
& $0.5690^{+0.0008}_{-0.0012}$ \\
& $\sigma$ ($10^{-3}$ keV)
& $2.4 \pm 0.6$
& $1.9^{+1.0}_{-0.8}$ \\
& Norm ($10^{-4}$)
& $5.3 \pm 1.7 $
& $7.8 \pm 1.8 $ \\

\hline
& $\chi^2$/dof
& $1167/983$
& $1123/979$ \\
\hline
\end{tabular}
\begin{tablenotes}
\footnotesize 
\item The quoted uncertainties are reported at the 90\% confidence level.
\item[$\dag$] Width tied to the corresponding narrow Fe K line.
\item[$\ddagger$] Additional Gaussian emission lines included only for the \\\textit{XMM-Newton/EPIC-pn} spectrum.
\item[*] Kept frozen during the fit.
\item[\#]
{\tt Model 1: constant*TBabs*2edge*(5gauss + bbodyrad + thcomp*bbodyrad)} \\
{\tt Model 2: constant*TBabs*2edge*(expabs*powerlaw + \\ relconv*rfxconv*thcomp*bbodyrad +  4gauss  + bbodyrad + thcomp*bbodyrad)}
\end{tablenotes}
\end{threeparttable}
}
\end{table}

\section{Polarimetric analysis}\label{Polarimetric Analysis}

\begin{figure}
    \centering
    \includegraphics[width=0.5\textwidth]{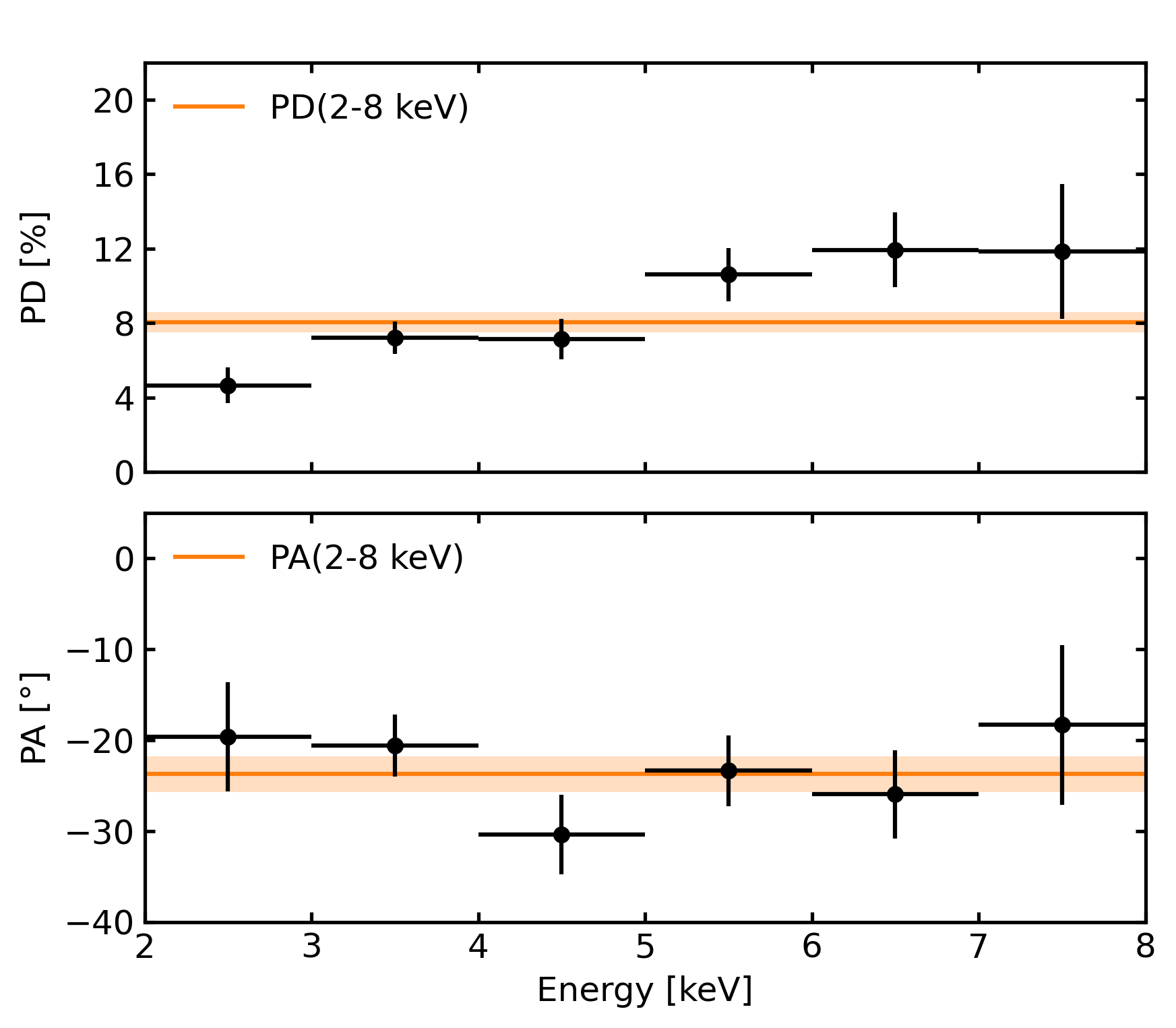}    \caption{Polarisation degree (top) and angle (bottom) versus energy in 1~keV bins obtained with the \texttt{ixpepolarization} task. Errors correspond to the 68\% confidence level. The orange line represents the average PD and PA over the 2--8 keV range, with the associated $1\sigma$ error highlighted by the shaded regions.}
    \label{fig:Pol.En}
\end{figure}

The X-ray polarisation of 4U 1822--37 can be derived in a model-independent way using the \texttt{ixpepolarization} task within HEASoft. This tool computes the Stokes parameters ($I$, $Q$, and $U$) and the PD and PA within a given spatial region and energy range. 
We detect significant polarisation in the 2--8 keV energy band, with 
${\rm PD}=7.9\pm0.6\%$ and ${\rm PA}=-24^\circ\pm2^\circ$. 
The measured PD is well above the 99\% minimum detectable polarisation, ${\rm MDP}_{99}=1.65\%$, corresponding to a nominal detection significance of ${\rm PD}/\sigma_{\rm PD}\simeq14.3$.
The polarisation derived in 1 keV-wide energy bins is reported in Table \ref{tab:Pol_flux} and shown in Fig. \ref{fig:Pol.En}: the PD appears to increase with energy, reaching $11\% \pm 4\%$ in the 7--8 keV bin, while the PA shows no significant rotation with energy. We also fitted the PD values with a linear model to test the significance of the increasing trend with energy. The resulting fit gives a $p$-value of 0.024, suggesting a possible increase in the PD at the 97.4\% confidence level. We applied the same test to the PA values and found a slope consistent with zero, confirming that there is no statistically significant rotation of the PA across the \textit{IXPE} band. Consistent polarisation values are also obtained using the \texttt{PCUBE} tool of \textsc{ixpeobssim} (v31.1.1; \citealt{Baldini2021}).

\begin{figure*}
    \centering
    \includegraphics[width=0.33\linewidth]{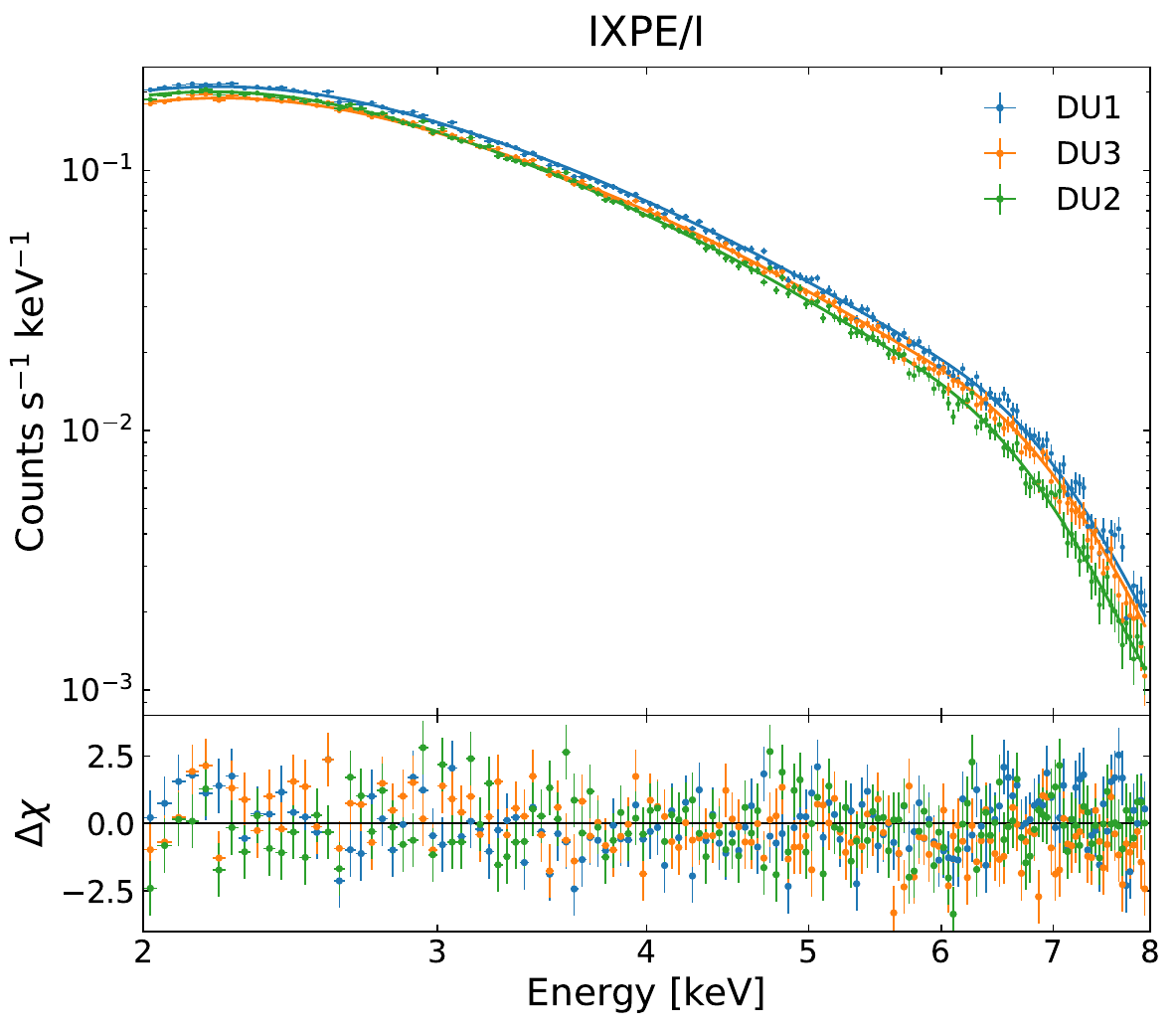}
    \includegraphics[width=0.33\linewidth]{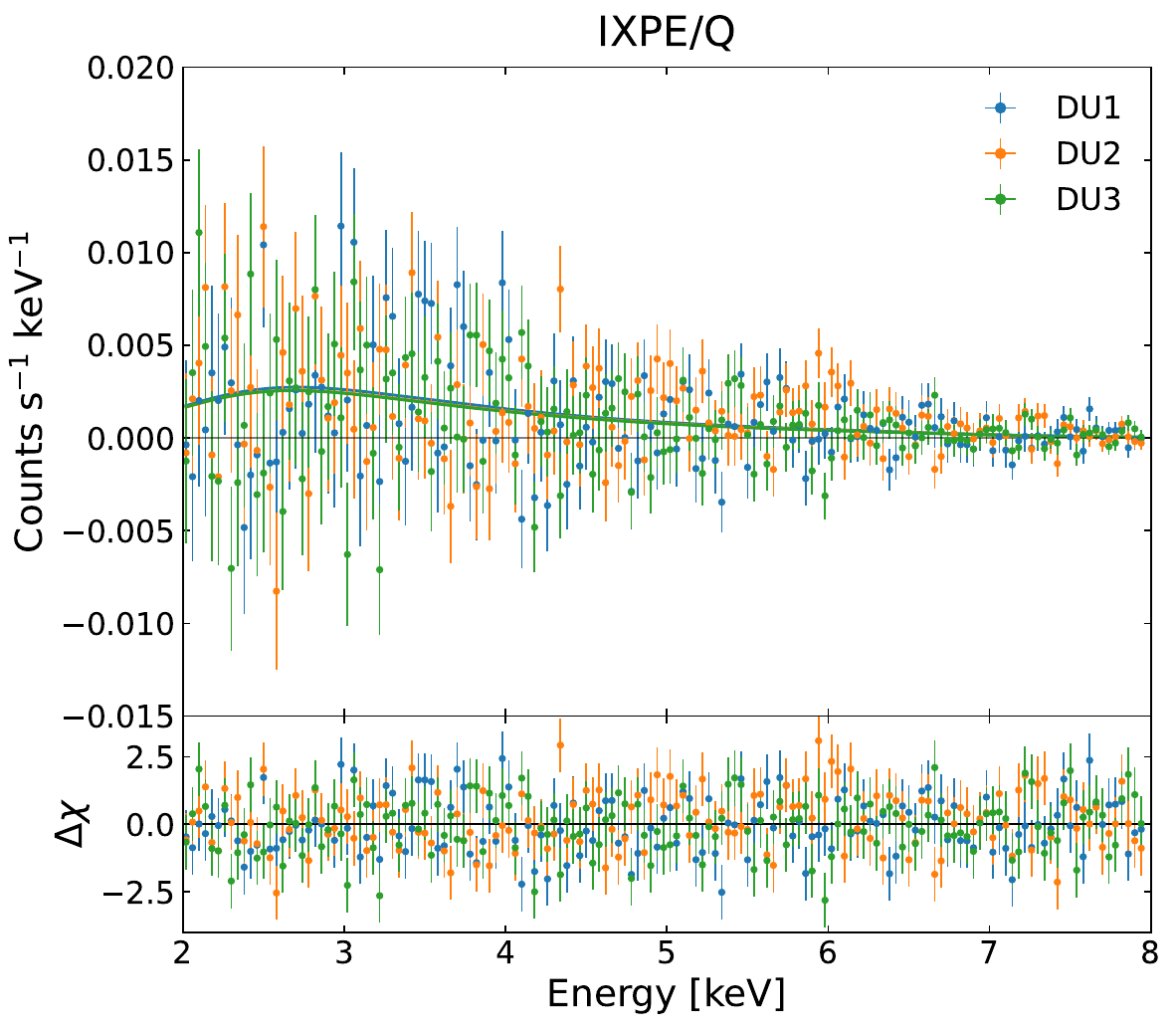}
    \includegraphics[width=0.33\linewidth]{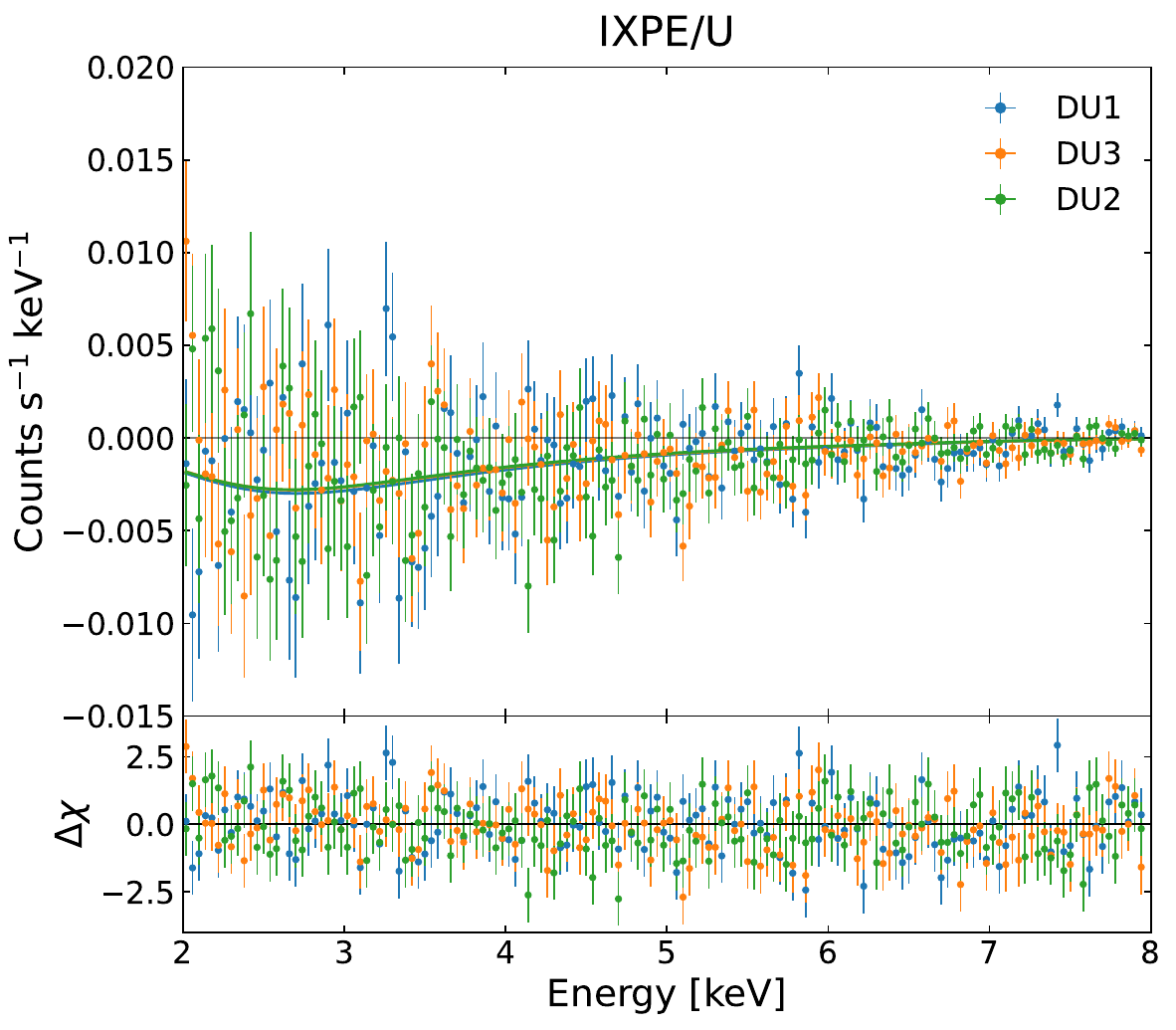}
    \caption{\textit{IXPE} $I$, $Q$, and $U$ Stokes spectra for each DU and the corresponding residuals in units of $\sigma$, obtained applying \texttt{polconst} to the best-fitting spectral model obtained with \textit{NuSTAR} and \textit{XMM-Newton.}}
    \label{fig:IXPE.Spectra}
\end{figure*}

We then performed a spectro-polarimetric analysis in \textsc{XSPEC}. Following common practice in \textit{IXPE} spectro-polarimetric analyses, we included a gain correction for the \textit{IXPE} spectra to account for small residual calibration and inter-calibration offsets with respect to the best-fit broadband spectral model \citep[e.g.][]{Gnarini.etAl.2024, Ursini_2024A&A...690A.200U}. For each DU, the gain parameters of the Stokes $Q$ and $U$ spectra were linked to those of the corresponding intensity spectrum, while the gain parameters and the cross-calibration constants were left free during the fit. All the other spectral parameters were fixed to the best-fit values obtained from the broadband spectral analysis.
We first applied the \texttt{polconst} model to the best-fit broadband spectral model, fixing all spectral parameters at their best-fit values, except for the cross-calibration constants. This configuration, hereafter \texttt{Model A}, yields $\chi^2/{\rm dof}=1411/1337$ and, in the 2--8 keV range, ${\rm PD}=6.8 \%\pm 0.5\%$ and ${\rm PA}=-24^\circ\pm2^\circ$, fully consistent with the model-independent measurement reported above. The corresponding \textit{IXPE} Stokes $I$, $Q$, and $U$ spectra and residuals are shown in Fig.~\ref{fig:IXPE.Spectra}.

As a next step, we replaced \texttt{polconst} with \texttt{pollin}, in which the PD and PA are allowed to vary linearly with energy:
\begin{equation*}
\begin{aligned}
    {\rm PD}(E)= {\rm PD}_{\rm 2\,keV} + (E-2\,{\rm keV})\times A_{\rm PD}., \\
    {\rm PA}(E)= {\rm PA}_{\rm 2\,keV} + (E-2\,{\rm keV})\times A_{\rm PA}.
\end{aligned}
\end{equation*}
After an initial fit, however, we found that the slope of the PA was $-1.8^\circ \pm 1.9^\circ$, fully consistent with zero. We therefore fixed the PA to be constant across all subsequent fits. Applying \texttt{pollin} to the total model (\texttt{Model B}) gave $\chi^2/{\rm dof}=1398/1336$, corresponding to a significant improvement over the constant polarisation case. An F--test gives $p=4.2\times10^{-4}$, corresponding to a preference at the $\sim$3.5--4$\sigma$ level for a linearly increasing total PD. The best-fit model gives ${\rm PD}_{2,{\rm keV}}=2.4\%\pm1.3\%$, increasing with a slope of $1.5\%\pm0.4\%$ per keV, and a PA of $-24^\circ\pm2^\circ$.

To investigate the contribution of individual spectral components to the net polarisation, we applied different polarisation models component-wise. In all these tests, the polarisation of the power-law component was tied to that of the Comptonised emission, since both are expected to arise from hybrid thermal or non-thermal Comptonisation. The soft thermal disc component was excluded from the spectro-polarimetric analysis because its contribution above 2~keV is negligible. We also tested models in which the reflection and Comptonised components were allowed to have independent PAs. However, this did not yield a qualitatively different solution, and the PA of the reflection component remained poorly constrained. By contrast, the PA of the dominant continuum component consistently lay between $-23^\circ$ and $-26^\circ$, very close to the total PA measured in the data. Since linking the two angles only marginally worsened the fit, and no significant PA rotation with energy was observed, we adopted a common PA as our reference configuration.

We first considered a model in which both reflection and Comptonisation were described by \texttt{polconst} (\texttt{Model C}). When left free, the reflection PD reached unphysical values, with a lower limit around $75\%$. We therefore fixed the reflection PD to 20\%, a value appropriate for a high-inclination system \citep[][]{Matt.1993,Podgorn_2025A&A...702A..43P}. With this choice, the Comptonised component is constrained to ${\rm PD}=6.3\%\pm0.5\%$ and ${\rm PA}=-26^\circ\pm 2^\circ$. Although the fit is acceptable ($\chi^2/{\rm dof}=1417/1337$), a constant polarisation for both components does not reproduce the observed increase in the PD with energy.

We then allowed both components to vary linearly with energy using \texttt{pollin} (\texttt{Model D}). Initially, we left the PD of the reflection component at 2 keV free to vary, but this parameter remained unconstrained. We therefore fixed it to 15\%, motivated by the high inclination of the source \citep{Podgorn_2025A&A...702A..43P}. This model improved the fit with respect to the constant component case, yielding $\chi^2/{\rm dof}=1403/1335$, but also revealed a strong degeneracy between the polarised components. In particular, when the reflection polarisation was allowed to increase with energy, the best-fit slope reached an upper limit of $7\%\,{\rm keV}^{-1}$, which implies a reflection PD of $\sim$ 60\% at 7--8~keV. 
For this reason, we explored more conservative mixed configurations. A model in which the reflection component was kept at a constant PD, while the Comptonised continuum was allowed to increase with energy (\texttt{Model E}), provided a comparably good fit ($\chi^2/{\rm dof}=1403/1336$) and yielded a more plausible behaviour of the individual components. In this case, the reflection PD was again fixed to 20\%, since the fit could not constrain it. We also explored the complementary case in which the reflection was allowed to vary with energy while the Comptonised component was kept constant (\texttt{Model F}). Formally, this model also reproduces the data, but drives the reflection polarisation to even more extreme values, reaching $\sim80$--90\% at 7--8 keV. Such values are well above those predicted by current theoretical calculations, suggesting that the present dataset does not robustly constrain the intrinsic energy dependence of the reflection component. The best-fit parameters of all the spectro-polarimetric models discussed above are reported in Table~\ref{tab:polmodels}. A physical interpretation of these different configurations is deferred to Section~\ref{Energy dependence of the polarisation}.

\begin{table*}[ht]
\centering
\renewcommand{\arraystretch}{1.2}
\setlength{\tabcolsep}{7pt}
\caption{Polarisation measured with \texttt{ixpepolarization} and percentage flux contribution of each spectral component according to the best-fit reflection model.}
\label{tab:Pol_flux}
\begin{tabular}{c|cc|cccc}
\hline
\multicolumn{1}{c|}{\multirow{2}{*}{Energy range}} 
& \multirow{2}{*}{PD ($\%$)} 
& \multirow{2}{*}{PA} 
& \multicolumn{4}{c}{Flux contribution} \\
\multicolumn{1}{c|}{} 
&  
&  
& Disc & Comp. & PL & Refl. \\
\hline
2--8 keV 
& $7.9\pm0.6$ 
& $-24^\circ\pm2^\circ$ 
& $<0.1\%$ & $88.5\%$ & $5.2\%$ & $5.3\%$ \\

2--3 keV 
& $4.6\pm1.1\%$ 
& $-18^\circ\pm6^\circ$ 
& $0.2\%$ & $88.6\%$ & $8.7\%$ & $2.2\%$ \\

3--4 keV 
& $7.1\pm0.9\%$ 
& $-21.^\circ\pm4^\circ$ 
& $<0.1\%$ & $91.3\%$ & $6.2\%$ & $2.2\%$ \\

4--5 keV 
& $7.0\pm1.1\%$ 
& $-30^\circ\pm5^\circ$ 
& $<0.1\%$ & $91.6\%$ & $5.2\%$ & $3.2\%$ \\

5--6 keV 
& $10.1\pm1.4\%$ 
& $-24^\circ\pm4^\circ$ 
& $<0.1\%$ & $89.9\%$ & $4.6\%$ & $5.5\%$ \\

6--7 keV 
& $11\pm2\%$ 
& $-26^\circ\pm5^\circ$ 
& $<0.1\%$ & $81.2\%$ & $3.9\%$ & $10.7\%$ \\

7--8 keV 
& $11\pm4\%$ 
& $-17^\circ\pm9^\circ$ 
& $<0.1\%$ & $90.4\%$ & $4.3\%$ & $5.4\%$ \\
\hline
\end{tabular}
\tablefoot{
The quoted polarisation uncertainties are at a confidence level of 68\%. 
The flux percentages were computed from the unabsorbed component fluxes in each energy band. }
\end{table*}

\begin{table}
\centering
\setlength{\tabcolsep}{5pt}
 \renewcommand{\arraystretch}{1.15}
\caption{Best-fit parameters of the spectro-polarimetric models discussed in the text.}
\label{tab:polmodels}
\begin{tabular}{ccccc}
\hline
Model & Parameters & \multicolumn{2}{c}{Value} & $\chi^2/{\rm dof}$ \\ \hline
\multirow{2}{*}{\texttt{Model A}} 
& PD (\%)      & \multicolumn{2}{c}{$6.8 \pm 0.5$}         & \multirow{2}{*}{1411/1337} \\
& PA $(^\circ)$ & \multicolumn{2}{c}{$-24 \pm 2$}         &                              \\ \hline
\multirow{3}{*}{\texttt{Model B}} 
& PD$_{\rm 2\,keV}$ (\%)   & \multicolumn{2}{c}{$2.4 \pm 1.3$}     & \multirow{3}{*}{1398/1336} \\
& $\alpha_{\rm PD}$ (\% keV$^{-1}$) & \multicolumn{2}{c}{$1.5 \pm 0.4$} &                              \\
& PA $(^\circ)$            & \multicolumn{2}{c}{$-24 \pm 2$}   &                              \\ \hline

& & Refl. & Compt. & \\ \hline

\multirow{2}{*}{\texttt{Model C}} 
& PD (\%)      & {[}20{]}                    & $6.3 \pm 0.5$ & \multirow{2}{*}{1417/1337} \\
& PA $(^\circ)$ & \multicolumn{2}{c}{$-26 \pm 2$}         &                              \\ \hline

\multirow{4}{*}{\texttt{Model D}} 
& PD$_{\rm 2\,keV}$ (\%)   & {[}15{]}                    & $<3.1$     & \multirow{4}{*}{1403/1335} \\
& $\alpha_{\rm PD}$ (\% keV$^{-1}$) & $< 7$   & $1.7^{+0.3}_{-0.8}$  &                              \\
& PA $(^\circ)$            & \multicolumn{2}{c}{$-27 \pm 2$}   &                              \\
 \hline
\multirow{3}{*}{\texttt{Model E}} 
& PD$_{\rm 2\,keV}$ (\%)   & {[}20{]}             & $1.4 \pm 1.3$           & \multirow{3}{*}{1403/1336} \\
& $\alpha_{\rm PD}$ (\% keV$^{-1}$)  & --                   & $1.7 \pm 0.5$         &                              \\
& PA $(^\circ)$                      & \multicolumn{2}{c}{$-27 \pm 2$} &                              \\ \hline

\multirow{4}{*}{\texttt{Model F}} 
& PD$_{\rm 2\,keV}$ (\%)   &  [15]          & $4.7 \pm 0.7$  & \multirow{4}{*}{1407/1336} \\
& $\alpha_{\rm PD}$ (\% keV$^{-1}$) & $13 \pm 4$  & --                      &                              \\
& PA $(^\circ)$            & \multicolumn{2}{c}{$-27 \pm 2$}   &                              \\
\hline
\end{tabular}
\tablefoot{
Values fixed during the fit are reported in square brackets. For Models C--F, the power-law polarisation is tied to that of the Comptonised component.}
\end{table}

\begin{table}
 \renewcommand{\arraystretch}{1.15}
\caption{Phase-resolved X-ray polarisation measured with the \texttt{ixpepolarization} task as a function of orbital phase.}\label{tab:Pol}      
\centering
\begin{tabular}{c c c}         
\hline\hline       
  Phase & PD (\%) & PA (deg) \\
\hline
0.00-0.15 & 5.5 $\pm$ 1.7 & -24 $\pm$ 9 \\
  0.15-0.36 & 8.0 $\pm$ 1.2 & -27 $\pm$ 4 \\
  0.36-0.56 & 9.9 $\pm$ 1.1 & -22 $\pm$ 3  \\
  0.56-0.76  & 8.1$\pm$ 1.2 & -22 $\pm$ 4 \\
  0.76-1.00 & 6.5 $\pm$ 1.2 & -25 $\pm$ 5 \\
\hline                                            
\end{tabular}
\tablefoot{Errors correspond to a confidence level of 68\%.}
\end{table}

\subsection{Phase-resolved analysis}
\begin{figure}
    \centering
    \includegraphics[width=0.5\textwidth]{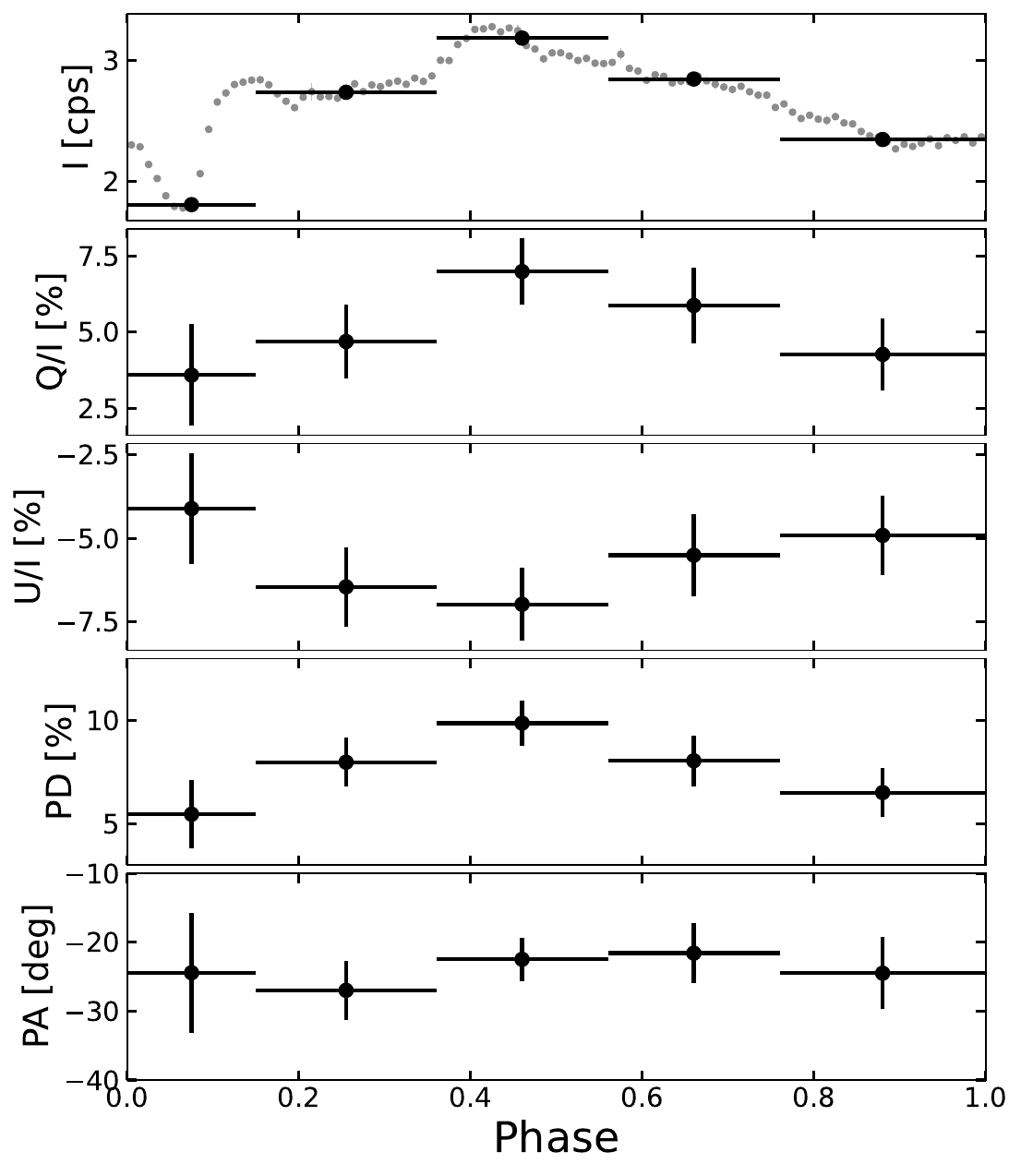}
    \caption{\textit{IXPE} count rate, normalised Stokes parameters, PD, and PA as a function of orbital phase in the 2--8~keV energy range. Errors correspond to the 68\% confidence level.}
    \label{fig:Pol.Phase}
\end{figure}

We performed a phase-dependent polarimetric analysis to determine whether the polarisation varies over the orbital period. 
We calculated the orbital period at the epoch of the \textit{IXPE} observation (MJD 60951), propagating the most recent ephemeris reported by \cite{Iaria_2024A&A...683A..79I}, obtaining $P_{\rm orb}=20054.39648$ s.  We obtained the folded profile by folding the background-subtracted data, summing the three DUs, and using 100 equally spaced bins (Fig. \ref{fig:Pol.Phase}). For phase-dependent polarimetric analysis, we adopted five phase intervals, corresponding to the eclipse ($\phi = 0-0.15$), the rise ($\phi = 0.15-0.36$), the peak ($\phi = 0.36-0.56$), and two bins during the decline ($\phi = 0.56-0.76,0.76-1$). We created the good time intervals (GTIs) for each phase interval and extracted the polarisation using the \texttt{ixpepolarization} task. Both the PD and the PA are well constrained in all orbital phase bins. As shown in Fig. \ref{fig:Pol.Phase}, the PD varies with orbital phase, reaching a maximum of 9.9\% $\pm$ 1.1\% at the phase corresponding to the highest source flux, and decreasing to 5.5 \% $\pm$ 1.7\% during eclipse. On the other hand, the PA remains consistent within the uncertainties throughout the orbital cycle. This behaviour suggests that the eclipse affects the relative contribution of the polarised emission components, although a more detailed interpretation is deferred to Sect. \ref{phase}.

Given that this source has repeatedly shown X-ray pulsations in previous studies, we conducted a pulse-phase-resolved polarimetric analysis to investigate the possible dependence of the polarisation properties on the NS spin phase.
Since the photon arrival times are modulated by the orbital motion of the binary system, we corrected the event times for the Römer delay assuming the orbital parameters reported in the literature. In particular, we adopted a projected semi-major axis of $a \sin i = 1.006(5)$~lt-s, as reported by \citet{jonker_01}. After applying the orbital correction, we searched for the spin signal using the \texttt{efsearch} task within the \textsc{XRONOS} package, adopting the start time of the observation as the reference epoch.
To obtain the most accurate estimate of the spin period at the epoch of our observation, we extrapolated the secular spin evolution reported by \cite{Iaria_2024A&A...683A..79I}, deriving an expected period of 0.590844360704 s. 
We searched for the spin signal with \texttt{efsearch} by scanning a narrow range of trial periods around the value predicted from the secular spin evolution, using different grid resolutions to verify the robustness of the result. No significant peak was detected in the resulting $\chi^2$ curves.
We interpret this non-detection as a consequence of both the limited photon statistics of the \textit{IXPE} observation and the intrinsically weak pulsation amplitude of the source. Indeed, previous studies reported a pulsed fraction below 0.5\% in the 2--6~keV band and below 1\% in the 6--10 keV band \citep{jonker_01, Iaria_2024A&A...683A..79I}. In this context, the relatively small effective area of \textit{IXPE}, especially at higher energies where the pulsation is expected to be stronger, prevents the detection of a statistically significant periodic signal.
As a further check, we folded the light curves at the expected spin period used in the timing search. 
The modulation is not statistically significant: the constant model gives $\chi^2$-test probability of $p=0.061$, while the sinusoidal fit gives a semi-amplitude pulsed fraction of $PF_{\rm sin}=1.3\pm0.4\%$. We therefore do not claim a detection of the spin modulation in the \textit{IXPE} data. As a conservative estimate, this corresponds to a 3$\sigma$ upper limit of $PF_{\rm sin}\lesssim2.7\%$, which is above the pulsed fractions previously reported for 4U~1822–37. We further restricted the analysis to energies above 7 keV, where a marginal modulation is apparent. However, once the data were divided into spin-phase bins, the counting statistics became too poor to allow for a meaningful extraction of the polarisation parameters in each bin. We therefore conclude that, for this source, a pulse-phase-resolved polarimetric analysis is currently not feasible with the available \textit{IXPE} statistics, and that significantly larger exposure times or a higher-throughput polarimetric mission would be required.

\section{Discussion}
We analysed the first simultaneous X-ray spectro-polarimetric observation of 4U 1822--37 obtained with \textit{IXPE} together with \textit{XMM-Newton}, \textit{Swift}, and \textit{NuSTAR}.
A model-independent analysis yields a band-averaged polarisation in the \textit{IXPE} 2 to 8 keV range of PD = 7.9 $\pm$ 0.6\% and PA = $-24^{\circ} \pm 2^\circ$. Such a high PD for a LMXB is exceptional. In the \textit{IXPE} literature, weakly magnetised NS-LMXBs generally show PD values below about five per cent, with the highest reported for XTE J1701--462 in the horizontal branch \citep{Cocchi_2023A&A...674L..10C}.
A similarly high PD has been reported only for 2S 0921--630, an ADC system with a nine-day orbit and an inclination comparable to that of 4U 1822--37 \citep{Tomaru_2026MNRAS.547ag498T}. The comparable PD measured in these two sources suggests that strong X-ray polarisation may be a common outcome of high-inclination, scattering-dominated ADC geometries, with very high inclination and scattering in an extended medium acting as the key drivers of the measured polarisation. In what follows, we first discuss the results of the spectral analysis alone, then combine them with the polarimetric constraints from the best-fit model to reconstruct the system geometry and show how the synergy between spectroscopy and polarimetry is decisive.

\subsection{Spectra decomposition}
The broadband spectrum is well described by a soft blackbody, a Comptonised component, a hard power-law tail, and a relativistically blurred reflection component. The blackbody marks a soft thermal excess at $kT_{\rm bb}=0.179^{+0.002}_{-0.003}$ keV that is plausibly associated with the accretion disc. The Comptonised spectrum arises from seed photons at $kT_{\rm seed}=1.351^{+0.006}_{-0.003}$ keV that are up-scattered by electrons with temperature $kT_{\rm e}=4.187^{+0.008}_{-0.007}$ keV. The \texttt{thcomp} covering fraction is $f_{\rm cov}=0.647^{+0.001}_{-0.009}$, implying that roughly 35\% of the seed-photon radiation reaches the observer without interacting with the corona. Consistent with the literature, this suggests that the Comptonising region is compact and does not completely surround the inner source. 
The Thomson optical depth, $\tau$, of the Comptonising corona can be calculated through the approximate analytical solution of the Kompaneets equation \citep[see][]{Zdziarski_1996MNRAS.283..193Z, Sunyaev_1980A&A....86..121S}.
In this framework, the photon index is related to the average number of scatterings experienced by a photon and can be written as 
\begin{equation}
\Gamma = \left[\frac{9}{4} + \frac{1}{\theta\,\tau\,\left(a_1 + a_2 \tau\right)}\right]^{1/2} - \frac{1}{2},    
\end{equation}
where $\theta = kT_{\rm e}/m_{\rm e}c^2$ is the dimensionless electron temperature, and the coefficients $a_1$ and $a_2$ account for the geometry of the scattering medium and higher-order corrections. Following \citep{Zdziarski_1996MNRAS.283..193Z, Zdziarski_2020MNRAS.492.5234Z}, we adopted $a_1 = 1.2$ and $a_2 = 0.25$, and assumed $\xi(\theta) \simeq 1$, which is appropriate in the mildly relativistic regime relevant for our measured temperatures ($kT_{\rm e} \sim 4.19 \, $keV).
By inverting the above relation, we derive an optically thick corona with $\tau = 20.8^{+0.2}_{-0.1}$.

The apparent size of the soft blackbody emitter follows from the \texttt{bbodyrad} normalisation, $K=R_{\rm km}^2/D_{10}^2$, where $R_{\rm km}$ is the source radius in km and $D_{10}$ is the distance in units of 10 kpc. Adopting a distance of 6.1~kpc \citep{Arnason_2021MNRAS.502.5455A}, the inferred apparent radius is $R_{\rm bb}=25.6^{+6.3}_{-0.5}$ km. In the same way, using the normalisation of the seed \texttt{bbodyrad} that feeds \texttt{thcomp}, we obtain an apparent seed-photon radius of $R_{\rm seed}=2.44\pm0.01$ km.

Following \citet{Anitra_2021A&A...654A.160A} and \citet{Iaria_2024A&A...683A..79I}, and using the independently inferred parameters for the system (magnetic field $B=(7.9\pm0.5)\times10^{10}$ G, neutron-star mass $M_{\rm NS}=1.61\,M_\odot$, radius $R_{\rm NS}=10$ km, and spin-period derivative $\dot P=-2.64(2)\times10^{-12}\ {\rm s\,s^{-1}}$), the intrinsic luminosity is expected to be L$_{\rm intr}\simeq1.5\times10^{38}\ {\rm erg\,s^{-1}}$. By contrast, our broadband fit yields an unabsorbed luminosity of only $L_{\rm obs}\simeq6.1\times10^{36}\ {\rm erg\,s^{-1}}$. This discrepancy is naturally explained if, at the very high inclination of the system, the direct inner emission is blocked by the outer disc rim and only a small fraction is scattered into the line of sight by an extended, optically thin accretion-disc corona. 
If we take $L_{\rm int}/L_{\rm obs }\approx 25$, the corona optical depth is  $\tau_{\rm C}\approx0.05$ (since $F_{\rm obs}\approx \tau_{\rm C} L_{\rm intr}/4\pi D^2$; \citealt{Iaria_13}). This picture therefore requires two distinct scattering regions, as already discussed in previous works: a compact, optically thick Comptonising region close to the NS, responsible for the observed Comptonised continuum and characterised here by $\tau\simeq 20.8$, and an extended, optically thin ADC with $\tau_{\rm c}\approx0.05$, which allows only a small fraction of the otherwise obscured inner emission to reach the observer through scattering.
A direct consequence is that the apparent normalisations of the thermal components do not trace the true emitting region.
Since the radius goes as $\sqrt{L}$, we should correct the radii by a factor of $\sqrt{L_{\rm intr}/L_{\rm obs}}=5$. As a consequence, the true blackbody and seed photon radii are  $R_{\rm bb}=127^{+31}_{-3}$ km and $R_{\rm seed}=12.09^{+0.02}_{-0.04}$ km.
In this picture, the soft blackbody is consistent with disc emission, while the seed-photon emitter is compatible with the surface of the NS.
This estimate is also consistent with the inner disc radius inferred from reflection. From the relativistic reflection fit we obtained $R_{\rm in}=59^{+7}_{-9} \, R_{\rm g}$. For a NS mass of $1.61\,M_{\odot}$, the gravitational radius is $R_{\rm g}=GM/c^{2}\approx 2.38$ km, which gives $R_{\rm in}\approx 140$~km. 
We stress that this radius refers to the broadband reflection component and it should therefore not be identified with the location of all the reprocessing material in the system. In particular, the recent XRISM Doppler-tomography study by \cite{Sameshima_2026PASJ...78..976S} localised the narrow neutral Fe K$\alpha$ fluorescence line in a region close to the accretion stream--disc overflow. This line traces a specific cold or mildly ionised fluorescence site, whereas in our modelling the narrow Fe-K features are treated separately from the self-consistent reflection component. The XRISM result therefore adds an important geometrical piece to the already complex picture of {\source}, but should not be confused with the broadband reflection component, which in our interpretation is associated with the inner accretion disc.

However, we must verify that these radii are consistent with a disc truncated by the NS magnetosphere. Given the presence of X-ray pulsations, the inner accretion flow is expected to be channelled along the magnetic field lines and the disc truncated at the magnetospheric radius, $R_{\rm m}$.
Following \citet{Sanna_17} we can write
\begin{equation}
R_{\rm m} = \phi\Big(\frac{\mu^{4}}{2GM\dot M^{2}}\Big)^{1/7},
\end{equation}
where $G$ is the gravitational constant, $M$ the NS mass, $\mu=BR_{\rm NS}^{3}$ the magnetic dipole moment, $\dot M$ the mass accretion rate, and $\phi$ a geometry factor that depends on the accretion--disc structure. 
For a Shakura-Sunyaev disc and fully ionised gas with mean molecular weight $\kappa=0.615$, $\phi$ is about 0.27 \citep{Shakura_1973A&A....24..337S}. As already reported in literature \citep[see][]{Anitra_2021A&A...654A.160A, Iaria_2024A&A...683A..79I}, assuming an intrinsic luminosity of $L_{\rm intr}\simeq1.5\times10^{38}\ {\rm erg \, s^{-1}}$ gives $R_{\rm m}\simeq105\pm5\ {\rm km}$.
This radius is smaller than both the blackbody radius and the inner disc radius from reflection, as expected if the magnetosphere truncates the disc while the regions that dominate the soft thermal emission and the reflection signatures lie slightly farther out. This is consistent with the scenario discussed by \citet{Iaria_2024A&A...683A..79I}.
It is worth noting that both the reflection inner disc radius and the magnetospheric radius are not affected by the presence of the optically thin scattering corona, since their estimates do not rely on the directly observed flux. 

\subsection{Energy dependence of the polarisation}\label{Energy dependence of the polarisation}
\begin{figure}
    \centering
\includegraphics[width=0.5\textwidth]{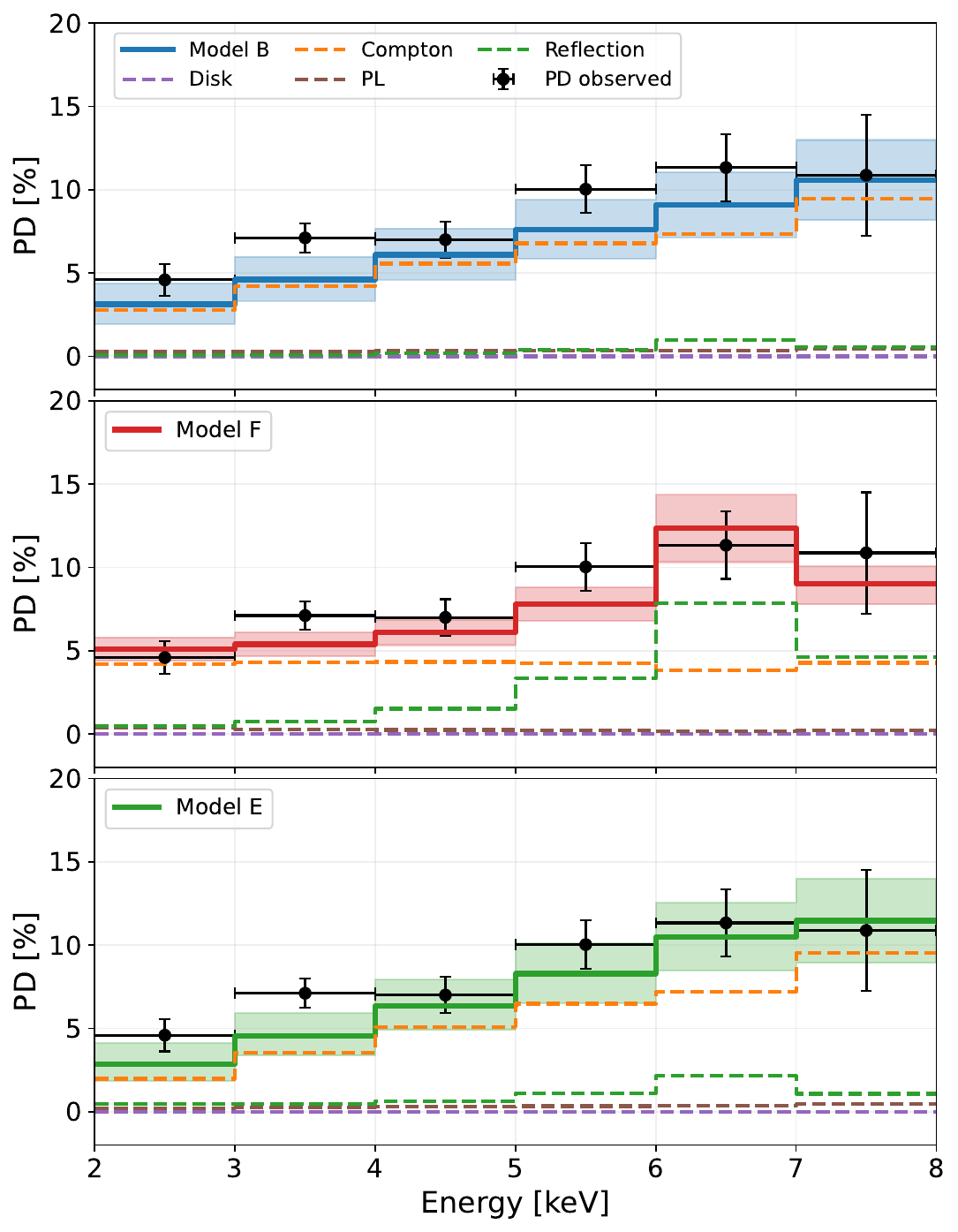}
\caption{Polarisation degree as a function of energy for three different models: \texttt{Model B} (top panel), \texttt{Model F} (middle panel), and \texttt{Model E} (bottom panel). In each panel, the solid line represents the total expected polarisation obtained from the Stokes-weighted combination of the individual spectral components. The shaded region indicates the 68\% confidence interval derived from Monte Carlo simulations, accounting for the uncertainties in both the polarisation parameters and the flux contributions of the different components. The dashed curves show the flux-weighted contributions of the individual spectral components, while the black points correspond to the values measured with \texttt{ixpepolarization}.}\label{fig:PDvspollin}
\end{figure}

The clearest polarimetric result of this work is the high PD measured in the \textit{IXPE} band. In addition, the data show evidence for an increase in the total PD with energy, while the PA remains consistent with a constant value within the uncertainties.
This behaviour is seen in both the model-independent analysis and the spectro-polarimetric fits.
In physical terms, it suggests that the projected geometry responsible for the net polarised signal remains broadly stable across the \textit{IXPE} band, while the net polarisation becomes progressively stronger at higher energies.
A rising PD with energy is broadly consistent with the emerging \textit{IXPE} picture of accreting neutron-star systems, where similar trends are frequently observed or hinted \citep{Ursini2024Galax..12...43U, Gnarini.etAl.2024, DiMarco_20205_https://doi.org/10.1002/asna.20240126}. However, the absolute PD measured in 4U~1822--37 is much higher than in most weakly magnetised NS-LMXBs, most likely because the observed radiation is already dominated by scattering in the extended ADC.
As discussed in Sect.~\ref{Polarimetric Analysis}, a more detailed decomposition of this trend among the individual spectral components is, however, not unique. In particular, we tested models in which the reflection and Comptonised components were allowed to have independent PAs, but this did not lead to a qualitatively different solution. The PA of the reflection component remained poorly constrained, whereas the PA of the dominant continuum component was always found to be very close to the total observed value. This, together with the lack of a significant PA rotation with energy, implies that the present data do not require two distinct projected symmetry axes in the \textit{IXPE} band. 

Using the best-fit parameters of the spectro-polarimetric models, we reconstructed the expected energy dependence of the total polarisation by combining the Stokes parameters of the individual spectral components and weighting each contribution by its relative flux in each energy bin. Fig. \ref{fig:PDvspollin} shows that different decompositions can reproduce the observed increase in the total PD, but also that the present data do not yet allow one to identify uniquely which component is primarily responsible for the trend. In particular, \texttt{Model E}, in which the reflection is kept at constant PD while the Comptonised component is allowed to increase with energy, provides a comparably good and more conservative description than models in which the reflection is forced to dominate the entire high-energy rise.

The role of the reflection component must therefore be treated with caution. Reflection is clearly required by broadband spectroscopy, but the current polarimetric data do not robustly constrain its intrinsic energy dependence. When the reflection PD is allowed to increase freely with energy, the fit tends to drive it towards very large values at high energies, reaching of order $\sim 70$--$90\%$ above 7~keV. 
Although recent calculations of the polarisation produced by reflected thermal emission \citep{Podgorny2026} show that highly ionised, optically thick material can become strongly polarised and can display an intrinsic increase in PD with energy in the \textit{IXPE} band, the very large values required by our free-reflection solution are difficult to interpret as a direct physical measurement. For the ionisation and seed-photon temperature inferred from our spectral fit, $\log\xi\simeq2.02$ and $kT_{\rm seed}\simeq1.35$~keV, the expected reflected PD is more likely of the order of a few to $\sim10$--20\%, depending on geometry. 
The $\sim60$--90\% values obtained by \texttt{Models D} and \texttt{F} above 7 keV are therefore more naturally interpreted as a symptom of model degeneracy. Thus, while the current dataset supports an increase in total polarisation with energy, it does not allow for a reliable determination of how this trend is split between the reflection and Comptonised continuum components.

\subsection{Orbital-phase dependence and eclipse behaviour}\label{phase}
The orbital-phase-resolved polarimetric analysis provides an additional and potentially important constraint on the geometry of 4U~1822--37. Although the statistical quality of the phase-selected \textit{IXPE} data remains limited, our results suggest that the PD decreases during eclipse. In contrast, the PA remains broadly consistent with the out-of-eclipse value (see Fig. \ref{fig:Pol.Phase}). This behaviour distinguishes 4U~1822--37 from the small but growing sample of high-inclination NS-LMXBs observed with \textit{IXPE}, in which dips or eclipses often increase the observed PD. 

In the eclipsing ADC source 2S~0921--630, \cite{Tomaru_2026MNRAS.547ag498T} reported ${\rm PD}=15\pm3\%$ during eclipse and ${\rm PD}=5.9\pm1.9\%$ outside eclipse, with no clear evidence for a corresponding change in PA between the two intervals. In AX~J1745.6--2901, \cite{Mikusincova_2026ApJ..1005..122M} found an even stronger increase, from about \(9\%\) outside eclipse to about \(34\%\) during dip or eclipse intervals, which they interpreted as the emergence of a scattering-dominated component when the central source is obscured. GX~13+1 also shows dip-related polarimetric variability, but with a more complex phenomenology: during dips the PD increases, and the PA exhibits swings of order \(\sim70^\circ\), indicating that the dip modulates not only the amplitude but also the dominant polarisation axis \citep{DiMarco2025ApJ...979L..47D}.

One possible explanation would be that during eclipse a weakly polarised soft component becomes relatively more important, thereby diluting the net polarised emission. However, the orbital spectroscopy of \citet{Iaria_13} already showed that, although the line emission varies with phase, the broadband continuum remains globally similar.
To verify that a spectral change does not simply cause the observed decrease in the PD, we extracted eclipse-selected spectra by selecting the time intervals corresponding to the eclipses in the \textit{NuSTAR} and \textit{EPIC-pn} light curves. The \textit{RGS} and \textit{Swift} data were not considered because of their lower statistical quality during the eclipse intervals. We then fitted the eclipse spectra with the same broadband model adopted for the phase-averaged analysis (\texttt{Model 2}). 
Owing to the lower statistics, some emission lines become only weakly constrained, and some discrete features are no longer significantly required. However, the broadband continuum above 2~keV remains globally similar to that observed out of eclipse, with no evidence for a major redistribution among the main continuum components. In particular, the soft blackbody remains negligible in the \textit{IXPE} band also during eclipse. This suggests that the decrease in the degree of polarisation is unlikely to be driven by an increased dilution from the thermal disc component.

The recent XRISM Doppler-tomography study of \citet{Sameshima_2026PASJ...78..976S} provides an additional comparison for the orbital-phase dependence observed with \textit{IXPE}. The authors showed that the centroid and profile of the narrow neutral Fe K$\alpha$ line vary with orbital phase, while the line-to-continuum ratio remains approximately constant. The decrease in the PD during eclipse is therefore unlikely to be driven by a modulation of the narrow Fe-line strength. This is also expected because the line contributes only a small fraction of the 2–8 keV flux, and the cold or mildly ionised material traced by it is not expected to provide a major polarising contribution through electron scattering. A significant scattering contribution would require a sufficiently ionised component with free electrons.
In this framework, the PD decrease is more naturally interpreted as a visibility effect, in which the companion star occults part of the extended ADC and changes the relative contribution of the visible scattering regions to the net polarised signal. 
This interpretation does not exclude the presence of additional ionised scattering material. Indeed, the narrow Fe XXVI feature shows that highly ionised gas is also present in the system. If this gas has a sufficiently large electron-scattering optical depth, it could in principle contribute to the polarised continuum, and its partial occultation during eclipse could affect the observed PD. However, the detection of this line alone does not allow us to quantify this effect or to link the ionised gas to the neutral Fe K$\alpha$ region identified by XRISM.

Such a visibility effect can arise if the eclipse acts directly on the extended scattering structure. In such a scenario, the eclipse does not enhance the relative contribution of a polarised scattered component by removing a weakly polarised central one; rather, it may occult the parts of the extended corona that, outside the eclipse, contribute most efficiently to the polarised signal. In the single-scattering Thomson regime, the linear polarisation depends on the scattering angle $\theta$ as
${\rm PD} = (1-\cos^{2}\theta)/(1+\cos^{2}\theta)$ , and is maximised for scattering angles close to $90^\circ$. Thus, if eclipse removes the regions that provide the most favourable scattering geometry towards the observer, the residual emission can naturally have a lower net PD while preserving a similar mean PA.

A related argument comes from scattering calculations in extended equatorial media. \cite{Nitindala_2025A&A...694A.230N} showed that, for an unpolarised central source scattered by an accretion-disc wind, the scattered PD is largest for low or intermediate wind opening angles, $\alpha_{\rm w}$. In their models, at small $\alpha_{\rm w}$ the incident radiation is mainly scattered in a plane perpendicular to the disc normal, producing a coherent positive polarisation signal, whereas for larger $\alpha_{\rm w}$ the scattering medium becomes more isotropic, scattering also occurs at higher latitudes, and competing polarisation contributions reduce the net PD. 
The total PD therefore reaches its maximum for an intermediate opening angle, around $\alpha_{\rm w}\sim20^\circ$, where the scattered fraction is sufficiently large, but the cancellation between different scattering planes is still limited.
This means that the net PD can decrease if the visible scattering region becomes more extended, more symmetric, or less favourably oriented.

A similar geometrical effect may occur during eclipse in 4U~1822--37. Since the system is viewed at very high inclination, the companion is expected to occult preferentially the low-latitude part of the extended corona, close to the orbital/disc plane. If this region provides the most efficient polarised scattering outside eclipse, the residual emission would be dominated by higher-latitude or more symmetric portions of the corona, with a lower polarimetric efficiency. This would naturally produce a lower PD during eclipse without requiring a major change in the broadband continuum.
A simple estimate shows that this scenario is geometrically plausible. The angular radius of the Roche-lobe filling companion as seen from the compact object is
$\delta \simeq \arcsin\left(R_{\rm L,2}/a\right)$,
where $R_{\rm L,2}$ is the Roche-lobe radius of the companion and $a$ is the binary separation. Using the approximation of \cite{Eggleton1983ApJ...268..368E},
\begin{equation}
\frac{R_{\rm L,2}}{a}=
\frac{0.49q^{2/3}}{0.6q^{2/3}+\ln(1+q^{1/3})},
\end{equation}
and the mass ratio, $q=M_2/M_1=0.24-0.27$, inferred by \cite{Munoz_05} for 4U~1822--37, we obtain $R_{\rm L,2}/a\simeq0.27$. This corresponds to $\delta\simeq16^\circ$. 
Therefore, although this estimate does not provide a full eclipse mapping of the extended corona, it shows that the companion can in principle occult a substantial low-latitude angular sector, comparable to the range of opening angles for which equatorial scattering models predict the highest polarimetric efficiency.

The observed decrease in the PD during eclipse can therefore be interpreted as eclipse selecting a less polarimetrically efficient portion of the extended corona. Rather than contradicting a scattering-dominated scenario, this behaviour supports the view that 4U~1822--37 is an extreme ADC system in which the observed X-ray emission is already dominated by scattered radiation outside eclipse.

\section{Conclusions}
We have presented the first X-ray spectro-polarimetric analysis of the eclipsing ADC source 4U~1822--37, based on a coordinated campaign with \textit{IXPE}, \textit{XMM-Newton}, \textit{NuSTAR}, and \textit{Swift}. The main results of this work can be summarised as follows:

\begin{itemize}
\item The broadband X-ray spectrum is well described by a soft thermal component, a Comptonised continuum, a hard power-law tail, and relativistically blurred reflection. These components are naturally interpreted within the ADC scenario. At the observed luminosity, $L_{\rm obs}\simeq6.1\times10^{36}\ {\rm erg\,s^{-1}}$, the apparent radii inferred are difficult to reconcile with the expected emitting regions. If the intrinsic luminosity is instead close to the Eddington limit and only a small fraction of the radiation is scattered into the line of sight by an extended, optically thin corona, the inferred radii become physically meaningful.

\item The model-independent \textit{IXPE} analysis yields a high 2--8~keV PD of ${\rm PD}=7.9\pm0.6\%$ and a PA of ${\rm PA}=-24^\circ\pm2^\circ$. This is among the highest PD values measured so far in a weakly magnetised NS-LMXB and is comparable only to the ADC source 2S~0921--630.

\item The PD increases with energy across the \textit{IXPE} band, while the PA remains consistent with a constant value. Spectro-polarimetric modelling shows that this behaviour can be reproduced by different decompositions among the Comptonised continuum and the reflection component. In particular, although the broadband spectroscopy clearly requires reflection, its intrinsic polarimetric behaviour is not robustly constrained by the present data. 

\item The orbital-phase-resolved \textit{IXPE} analysis suggests that the PD decreases during eclipse, from values of close to $\sim8$--10\% outside an eclipse to ${\rm PD}=5.5\pm1.7\%$ during an eclipse, while the PA remains broadly consistent with the phase-averaged value. This behaviour differs from other high-inclination NS-LMXBs observed with \textit{IXPE}, where dips or eclipses often increase the observed PD. The decrease in the PD is unlikely to be driven by enhanced dilution from the soft thermal component and is more naturally interpreted as an eclipse selecting a less polarimetrically efficient portion of the extended scattering structure.
\end{itemize}

Taken together, spectroscopy and polarimetry provide a consistent picture in which 4U~1822--37 is observed in an extreme high-inclination, scattering-dominated regime. The high band-averaged PD, the absence of a significant PA rotation, and the increase in the total PD with energy are compatible if the \textit{IXPE}-band emission is dominated by radiation processed in a geometrically coherent, extended ADC.
The decrease in PD during the eclipse further strengthens this interpretation. Rather than contradicting a scattering-dominated scenario, this behaviour suggests that the eclipse acts directly on the visible scattering structure, selecting a region with lower polarimetric efficiency. In this picture, the ADC is not merely an additional spectral component, but a central element that shapes the observed X-ray emission and its polarisation properties in 4U~1822--37.

\begin{acknowledgements}
      Part of this work was supported by \emph{ESO}, project
      number Ts~17/2--1.

AG was supported by an appointment to the NASA Postdoctoral Program at the Marshall Space Flight Center (MSFC), administered by Oak Ridge Associated Universities under contract with NASA. 
AM acknowledges support from the Fund Vera Rubin/Chile 2024, under the project DIA 1736 "Silent black holes around red supergiants", the H2020 ERC Consolidator Grant “MAGNESIA” under grant agreement No. 817661 (PI: Rea) and the National Spanish grant PID2023-153099NA-I00 (PI: Coti Zelati). 
FB acknowledges support from INAF Large Grant 2023 BLOSSOM F.O. 1.05.23.01.13. FC, SF and AT acknowledge financial support by the Istituto Nazionale di Astrofisica (INAF) grant 1.05.24.02.04: ``A multi frequency spectro-polarimetric campaign to explore spin and geometry in Low Mass X-ray Binaries''.
This work reports observations obtained with the Imaging X-ray Polarimetry Explorer (IXPE), a joint US (NASA) and Italian (ASI) mission, led by MSFC. The research uses data products provided by the IXPE Science Operations Center (MSFC), using algorithms developed by the IXPE Collaboration (MSFC, Istituto Nazionale di Astrofisica - INAF, Istituto Nazionale di Fisica Nucleare - INFN, ASI Space Science Data Center - SSDC), and distributed by the High-Energy Astrophysics Science Archive Research Center (HEASARC). This research has made use of data from the NuSTAR mission, a project led by the California Institute of Technology, managed by the Jet Propulsion Laboratory, and funded by NASA. Data analysis was performed using the NuSTAR Data Analysis Software (NuSTARDAS), jointly developed by the ASI Science Data Center (SSDC, Italy) and the California Institute of Technology (USA).
\end{acknowledgements}

\bibliographystyle{aa}
\bibliography{bliblio}

\begin{appendix}
    
\section{EPIC-pn energy-scale check and correction}\label{epic_shift}
\begin{figure}[h!]
    \centering
    \includegraphics[width=0.5\textwidth]{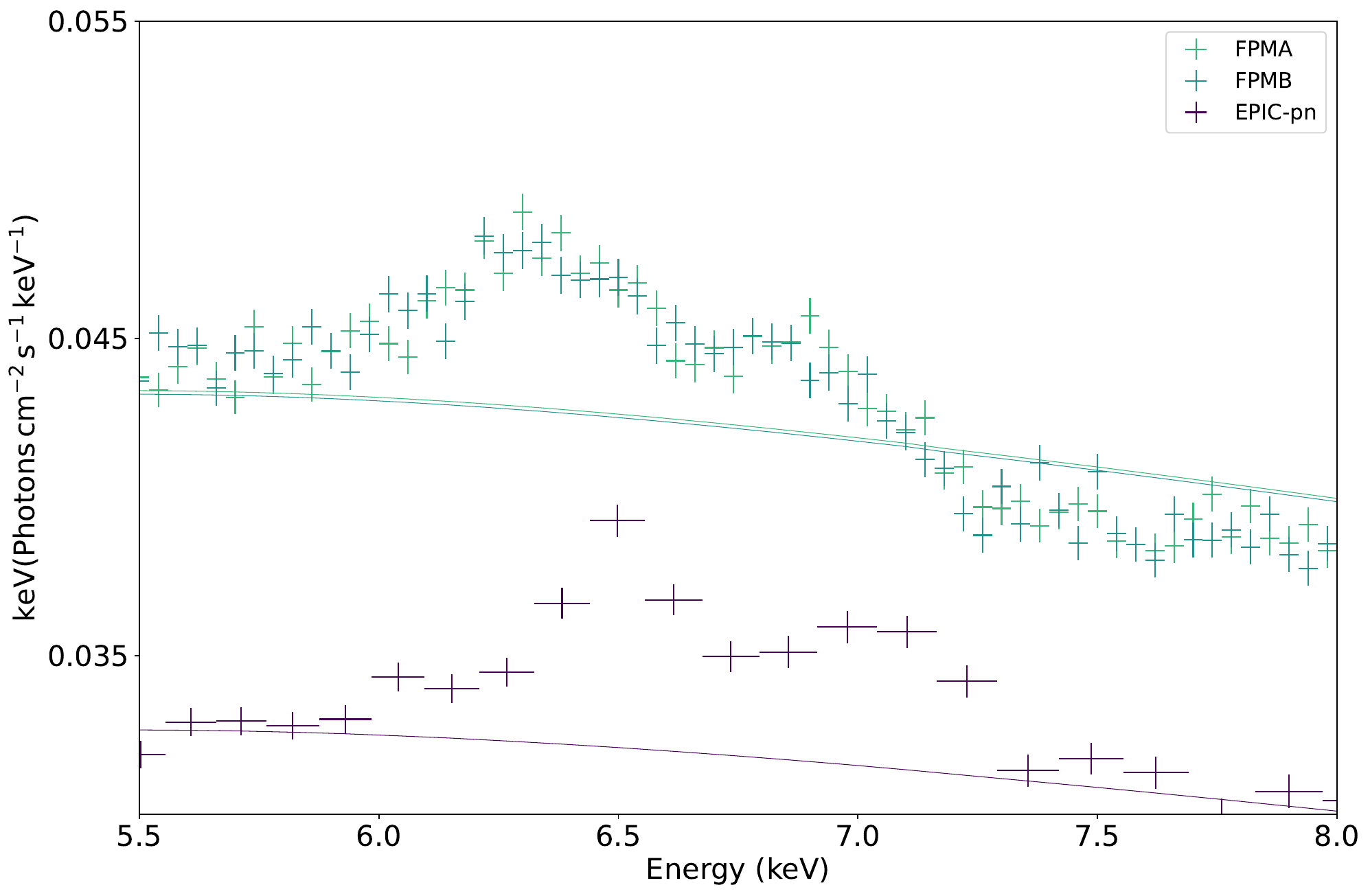}    \caption{Unfolded spectra from \textit{EPIC-pn} and \textit{NuSTAR} FPMA/FPMB in the iron-line region, fitted with a simple power law.}
    \label{comparision_lines}
\end{figure}

Bright sources observed in \textit{EPIC-pn} Timing mode may exhibit count rate dependent energy scale shifts. Although the standard processing applies a generic correction, in some cases it is not enough. 
The spectra we initially extracted show several line features in the 0.5--10 keV band at higher energies than expected, most notably the fluorescent Fe K$\alpha$ from neutral iron and the H-like Fe XXVI line. Fig. \ref{comparision_lines} displays the \textit{NuSTAR} and \textit{EPIC-pn} spectra in the iron-line region fitted with a simple power law, where the Fe K$\alpha$ centroid is found at $>6.53$~keV, whereas it is expected at $\sim6.4$~keV, and the Fe XXVI Ly$\alpha$ centroid is at $\sim7.2$~keV instead of the expected $\sim6.97$~keV.

A practical way to verify that the extracted spectrum is affected by count-rate--dependent energy shifts, and not by some physical effect such as a Doppler shifts from winds \citep[e.g.][]{Diaz_2016AN....337..368D,Ponti2012MNRAS.422L..11P}, is to focus on narrow windows around instrumental edges. The most useful are the Si K edge at about 1.84 keV, the Au M edge near 2.2 keV, and the Au L edge around 11.9 keV \citep[see ][]{CAL_SRN_0369}. If residuals appear systematically on both sides of these edges, or if the apparent edge energies are offset by the same amount in eV at low and high energies, the issue is an energy-scale shift. To quantify the shift in energy scale needed to minimise these residuals, we used the \textsc{XSPEC} command \texttt{gain fit}. To correct the effect, we applied the SAS task \texttt{evenergyshift} to create a new events file with the derived energy shifts, from which we then re-extracted the source and background spectra using the same regions as above.
\end{appendix}
\end{document}